\documentclass[preprint,12pt]{elsarticle}
\usepackage{graphicx}
\usepackage{dcolumn}
\usepackage{bm}
\usepackage{amsmath,amssymb,latexsym}
\usepackage{xspace}
\usepackage{hyperref}

\usepackage{booktabs} 
\usepackage{xcolor} 
\usepackage[expansion=false]{microtype} 
\usepackage{placeins} 
 
\newcommand{\pvec}[1]{\vec{#1}\mkern2mu\vphantom{#1}}

\hypersetup{hidelinks}
\begin{document}

\begin{frontmatter}

\journal{Computer Physics Communications}

\title{Multiscale Numerical Modelling of Ultrafast Laser--Matter Interactions: Maxwell--Two-Temperature Model--Molecular Dynamics (M-TTM-MD)}

\author[inst1]{O. Benhayoun\corref{cor1}}
\ead{obenhayoun@uni-kassel.de}

\author[inst1]{M. E. Garcia}

\cortext[cor1]{Corresponding author}

\address[inst1]{Institute of Theoretical Physics II and Center for Interdisciplinary Nanostructure Science and Technology (CINSaT), Universität Kassel, Heinrich-Plett-Straße 40, 34132 Kassel, Germany}

\begin{abstract}
In this work, we present a comprehensive numerical framework that couples numerical solutions of Maxwell’s equations using the Finite-Difference Time-Domain (FDTD) approach, Molecular Dynamics (MD), and the Two-Temperature Model (TTM) to describe ultrafast laser--matter interactions in metallic systems at the atomic scale. The proposed Maxwell-Two-Temperature Model-Molecular Dynamics (M-TTM-MD) bridges the gap between electromagnetic field propagation, electron--phonon energy exchange, and atomic motion, allowing for a self-consistent treatment of energy absorption, transport, and structural response within a unified simulation environment. The calculated electromagnetic fields incorporate dispersive dielectric properties that are represented using the Auxiliary Differential Equation (ADE) technique, while the electronic and lattice subsystems are dynamically coupled through spatially and temporally resolved energy exchange terms. The changes in the material topography are then reflected in the updated grid for the FDTD scheme. {We present additional simulations that compare FDTD and Beer--Lambert heating and the resulting material response in a gold sample with a groove, and examine wavelength-dependent near fields above a rough gold surface.}\ The developed M-TTM-MD model provides a self-consistent numerical framework that offers insights into laser-induced phenomena in metals, including energy transport and surface dynamics under extreme nonequilibrium conditions.
\end{abstract}


\end{frontmatter}

\section{Introduction}
Ultrafast laser--matter interaction has emerged as a key research area for understanding and controlling physical processes on femtosecond timescales. When a femtosecond laser pulse irradiates a metallic target, the absorbed energy is first deposited into the electronic subsystem, leading to a rapid increase of its temperature while the lattice remains at room temperature. The energy is then transferred
to the ions via electron--phonon coupling, such that both systems eventually reach thermal equilibrium. This nonequilibrium process can be crucial for a wide range of ultrafast phenomena that require ablation or surface modification of the target material. One important application is the formation of Laser-Induced Periodic Surface Structures (LIPSS), in which periodic and uniform ripples can be formed at the surface of a wide variety of materials, ranging from dielectrics to metals \cite{birnbaum1965semiconductor,benhayoun2021theory,bonse2009role,bonse2020maxwell,gurevich2016mechanisms,mastellone2022lipss,schwarz2018homogeneous,sipe1983laser,blumenstein2020formation,terekhin2019influence}.

To describe these dynamics, the Two-Temperature Model (TTM) \cite{anisimov1997theory,singh2010two} has been widely employed, providing a macroscopic picture of the coupled electron and lattice temperature evolution. However, this approach lacks atomistic resolution and cannot capture local structural transformations. Molecular Dynamics (MD) simulations, on the other hand, offer atomic-scale insight into the lattice response but generally rely on empirical or uniform heating schemes that neglect spatially varying energy absorption and electromagnetic effects.

To address these limitations, more recent studies have coupled MD with the TTM formalism \cite{ivanov2003combined,ivanov2019numerical,terekhin2022key}, yielding a more realistic description of lattice dynamics and phase transitions. However,   {many implementations prescribe an analytical optical source term, often laterally uniform for a uniformly illuminated planar target. Such sources can include depth attenuation and temperature-dependent optical coefficients, but they do not incorporate field interference, diffraction, or geometry-dependent electromagnetic redistribution} . Surface topography, for example, scatters the incident laser and can excite surface waves such as surface plasmon polaritons \cite{bonse2020maxwell,sipe1983laser,benhayoun2021theory,terekhin2019influence}, which significantly affect energy absorption. Moreover, multiple laser pulses progressively reshape the surface; at sufficiently high fluence, this can drive phase transitions, induce surface deformation, or even remove material via spallation or evaporation. Capturing this interpulse feedback is therefore essential for an accurate description of the sample’s temporal evolution and final resolidified structure.

  \begin{figure}[htbp]
     \centering
    \includegraphics[width=.75\linewidth]{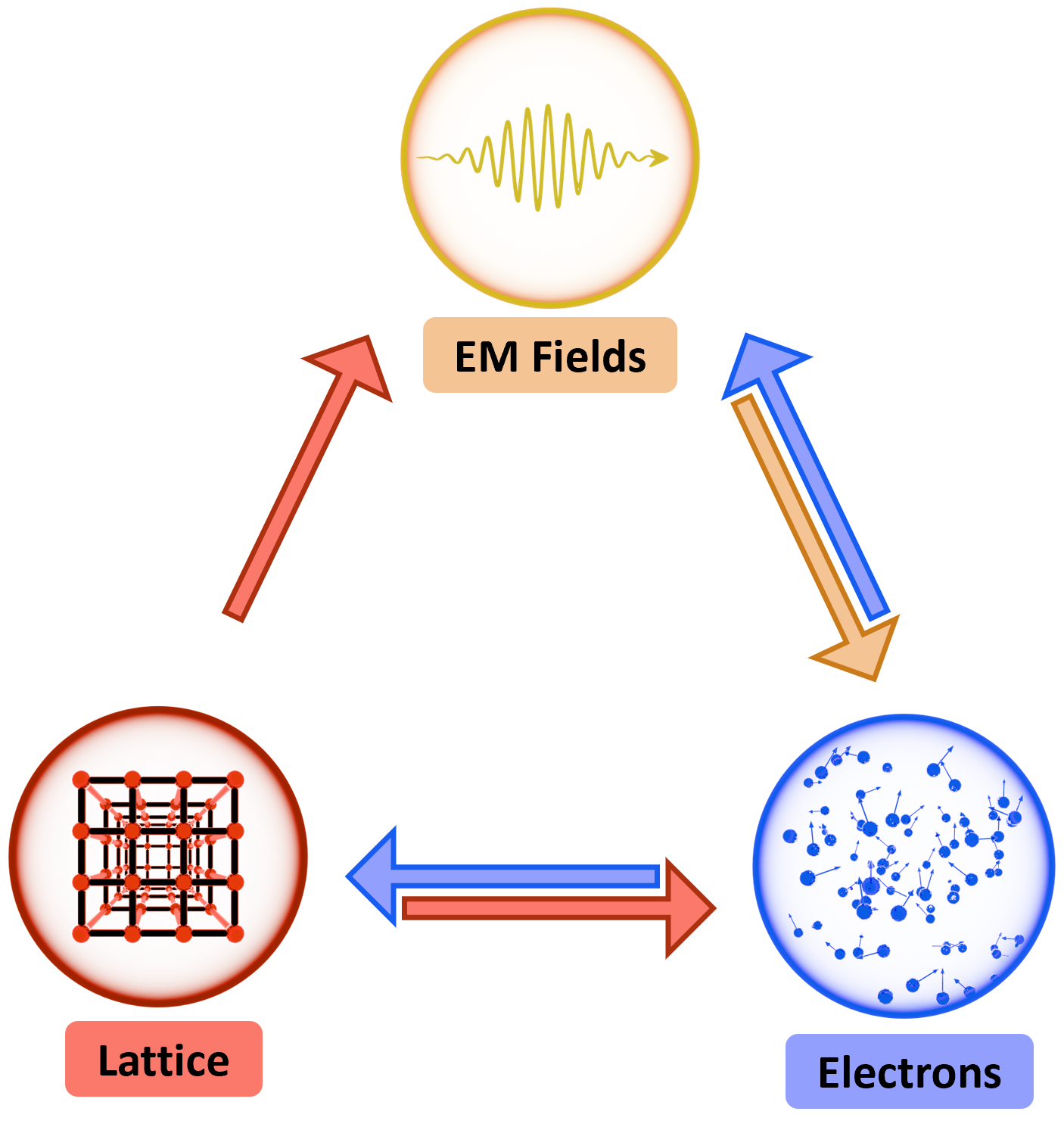}
    
     \caption{Schematic representation of the different interactions accounted for in the M-TTM-MD model. The electromagnetic (EM) wave propagates through the material, where it is absorbed by the electronic subsystem, {leading to a rapid increase} of the electron temperature $T_e$ as is represented by the yellow arrow from the EM Fields to the Electrons. The absorbed energy is then transferred to the lattice subsystem via electron-phonon coupling, raising the lattice temperature $T_l$ represented by the red and blue arrows connecting the Electrons and the Lattice. The elevated lattice temperature induces atomic motion and structural and surface changes, which in turn affect the material's properties and {modify the scattering and absorption of subsequent EM waves}, shown by the red and blue arrows from the Electrons and the Lattice to the EM Fields. This feedback loop continues throughout the laser--matter interaction process.}
    \label{fig:MTTMMD-interactions}
\end{figure}

In this work, we present a fully coupled Maxwell-Two-Temperature Model-Molecular Dynamics (M-TTM-MD) framework in which electromagnetic field propagation, nonequilibrium electron and lattice heating, and atomistic structural dynamics are treated self-consistently within a single computational model. The electromagnetic problem is solved using the Finite-Difference Time-Domain (FDTD) method \cite{yee1966numerical,taflove1993computational}, yielding a spatially and temporally resolved field distribution from which the locally absorbed energy density is evaluated and introduced as a source term for the electronic subsystem. The different interactions between the different subsystems can be summarized in Fig.~\ref{fig:MTTMMD-interactions}, where the absorption of the incident EM wave rapidly elevates the electron temperature $T_e$, after which electron-phonon coupling transfers that energy to the lattice and raises the lattice temperature $T_l$ on a picosecond timescale. The associated temperature elevation induces a rapid atomic motion, stress build-up, and phase transformations, thereby modifying the surface morphology. Variations in the material density as well as in the electron and lattice temperatures alter the modeled material's optical response at the interface and, consequently, the subsequent field surface distribution and absorption. This feedback between the lattice's thermodynamic properties (temperature, pressure and density), the electron's thermal diffusion equation and the electromagnetic fields' distribution forms a closed loop that ensures a self-consistent evolution of the coupled modules. The implementation targets large-scale computations through MPI-based parallelization and employs explicit energy accounting to maintain consistency across the FDTD, TTM, and MD components. The resulting framework enables quantitative, spatially resolved investigations of ultrafast energy transport and structural evolution in different materials under strongly nonequilibrium excitation.

\section{The \textrm{M-TTM-MD} model} \label{theory}

\subsection{Overview of the simulation framework}

\begin{figure}[htbp]
     \centering
      {\includegraphics[width=\linewidth]{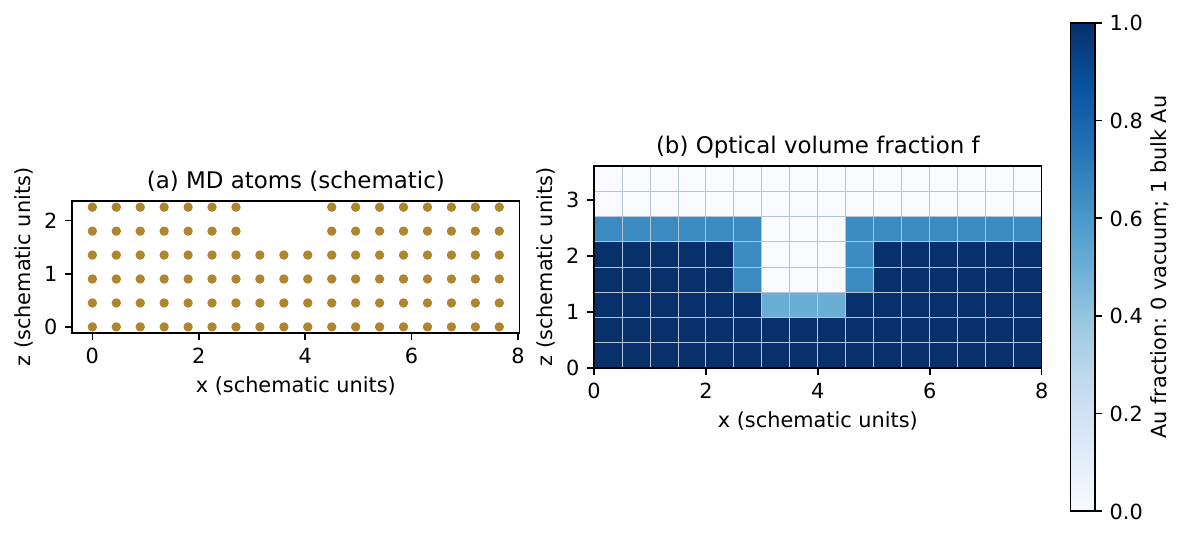}
    } \caption{Schematic   {atom-to-optical mapping. The left panel shows atomic positions; } the   {right panel shows } the   {bounded material fraction assigned to } continuum cells.   {White denotes vacuum ($f_{\rm Au}=0$)} ,   {dark blue denotes bulk occupancy ($f_{\rm Au}=1$), and intermediate colors denote partial occupancy. The color is a geometric fraction, not electron temperature or complex permittivity. This drawing illustrates } the   {mapping and is not a simulation snapshot} .}
    \label{fig:TTM-FDTD}
\end{figure}

As mentioned previously, the main objective of the proposed Maxwell-Two-Temperature Model-Molecular Dynamics (M-TTM-MD) is to derive, with high precision, a self-consistent source term for ultrafast laser excitation capable of interacting with metallic systems of varying geometrical complexity.
The Finite-Difference Time-Domain (FDTD) method is therefore employed to calculate the electromagnetic energy absorbed by the material, where the field is discretized on a uniform Yee grid \cite{yee1966numerical,taflove1993computational,schneider2010understanding}. Each cell stores the field components and is updated at every timestep to numerically solve Maxwell’s equations. In parallel, the TTM-MD framework simulates the atomic-scale dynamics of solids under external excitation by integrating Newton’s equations of motion while ensuring local energy balance between the electronic and lattice subsystems.
The integration of these two solvers requires the consistent coupling of their respective domains. The M-TTM-MD framework thus employs three spatial grids FDTD, TTM and MD. The MD (\textit{reference mesh}) consists of the discrete atomic positions, while the FDTD and TTM meshes are \textit{continuum meshes} that calculate the fields and temperatures of the different subsystems respectively, while updating spatially averaged quantities such as the pressure and material density from the MD grid. The resolutions and timesteps must be chosen to   {resolve the geometry and satisfy the stability conditions of each module. A uniform Yee grid with grid spacings $\Delta x$, $\Delta y$, and $\Delta z$ is chosen independently of the TTM and MD spacings. The source term transfer between the FDTD and the TTM grids uses cell-volume weights, and therefore, does not require an integer ratio between the two meshes, 
(} Fig.~\ref{fig:TTM-FDTD}).   {A practical starting point is
}\begin{equation}{\label{eq:fdtd-grid-size}
\Delta \alpha \lesssim \min\left(\frac{d_{\min}}{N_g},\frac{\lambda_{\min}}{N_\lambda},\frac{\delta_{\rm skin}}{N_s}\right),
}\end{equation} 
where $\alpha = x, y, \text{ or } z$ and   {$N_g$, $N_\lambda$, and $N_s$ denote the chosen numbers of cells across a geometrical feature, wavelength, and optical penetration depth. These choices provide an initial mesh estimate that resolves the smallest geometrical features, the shortest relevant wavelength within the computational domain, and the minimum optical skin depth, while limiting spatial discretization errors in the FDTD solution.
}

{The vacuum Yee CFL bound is
}\begin{equation}{\label{eq:cfl}
c\,\Delta t_\text{FDTD}\sqrt{\Delta x^{-2}+\Delta y^{-2}+\Delta z^{-2}}\leq 1.
}\end{equation}
{The material ADE recurrence and the thermal update impose additional restrictions. FDTD, TTM, and MD steps are synchronized in physical time.}   {The FDTD calculation continues beyond the temporal width of the incident laser intensity so that the residual stored electromagnetic and polarization energy becomes negligible and the cumulative absorbed, reflected, and transmitted energies stop changing appreciably. For an FDTD time step $\Delta t_\text{FDTD}$, the required number of steps is $N_F = \lceil t_{\rm opt}/\Delta t_\text{FDTD} \rceil$ where $t_{\rm opt}$ is the optical integration time.
Waves leaving the domain along nonperiodic directions are absorbed by CPML layers \cite{roden2000convolution}. Their thickness and separation from the sample are chosen to ensure that the absorbing boundaries do not significantly affect the calculated response.}

{The atomic positions are first mapped onto a cell-centered grid to determine the local gold fraction, $0\leq f_{\rm Au}\leq1$. A value of zero represents vacuum, whereas a value of one represents a region fully occupied by gold. These fractions are then transferred to the staggered locations of the electric-field components. At locations that are only partially occupied by gold, the
effective optical parameters are calculated as
}\begin{equation}{\label{eq:optical-fraction}
\epsilon_{\infty,\mathrm{eff}}=1+f_{\rm Au}(\epsilon_{\infty,\mathrm{Au}}-1),\qquad
\omega_{p,\mathrm{eff}}^2=f_{\rm Au}\omega_{p,\mathrm{Au}}^2,\qquad
a_{j,\mathrm{eff}}=f_{\rm Au}a_{j,\mathrm{Au}}.
}\end{equation}
{Here, $a_j$ denotes the strength of the $j$-th Lorentz oscillator. The optical parameters of bulk gold are evaluated from the material model at the selected electron and lattice temperatures. The local gold fraction then scales the effective coefficients through Eq.~\eqref{eq:optical-fraction}. This formulation uses linear mixing for the background permittivity and interpolation of temperature-dependent tables.
}
  {For gold at visible wavelengths, the implementation supports a temperature-dependent Drude--Lorentz response based on the material model of Ref.~\mbox{
\cite{ndione2024adaptive}}\hskip0pt
. The dielectric function of bulk gold is written as
}\begin{equation}{\label{eq:bulk-dielectric}
\epsilon_{\rm Au}(\omega;T_e,T_l)=\epsilon_\infty
-\frac{\omega_p^2}{\omega^2+i\gamma_D\omega}
+\sum_j\frac{a_j}{\omega_j^2-\omega^2-i\gamma_j\omega},
}\end{equation}
{where the damping rates are positive and the frequency-domain response uses the $e^{-i\omega t}$ convention. The fitted optical parameters are obtained by interpolating temperature-dependent tables. The time-dependent auxiliary differential equations (ADE) describe the evolution of the polarization and associated currents. The coefficients are refreshed at the prescribed material-update synchronization points, and the energy associated with changes in these coefficients is included in the energy balance.

}\begin{figure}[htbp]
    \centering
    \includegraphics[width=.75\linewidth]{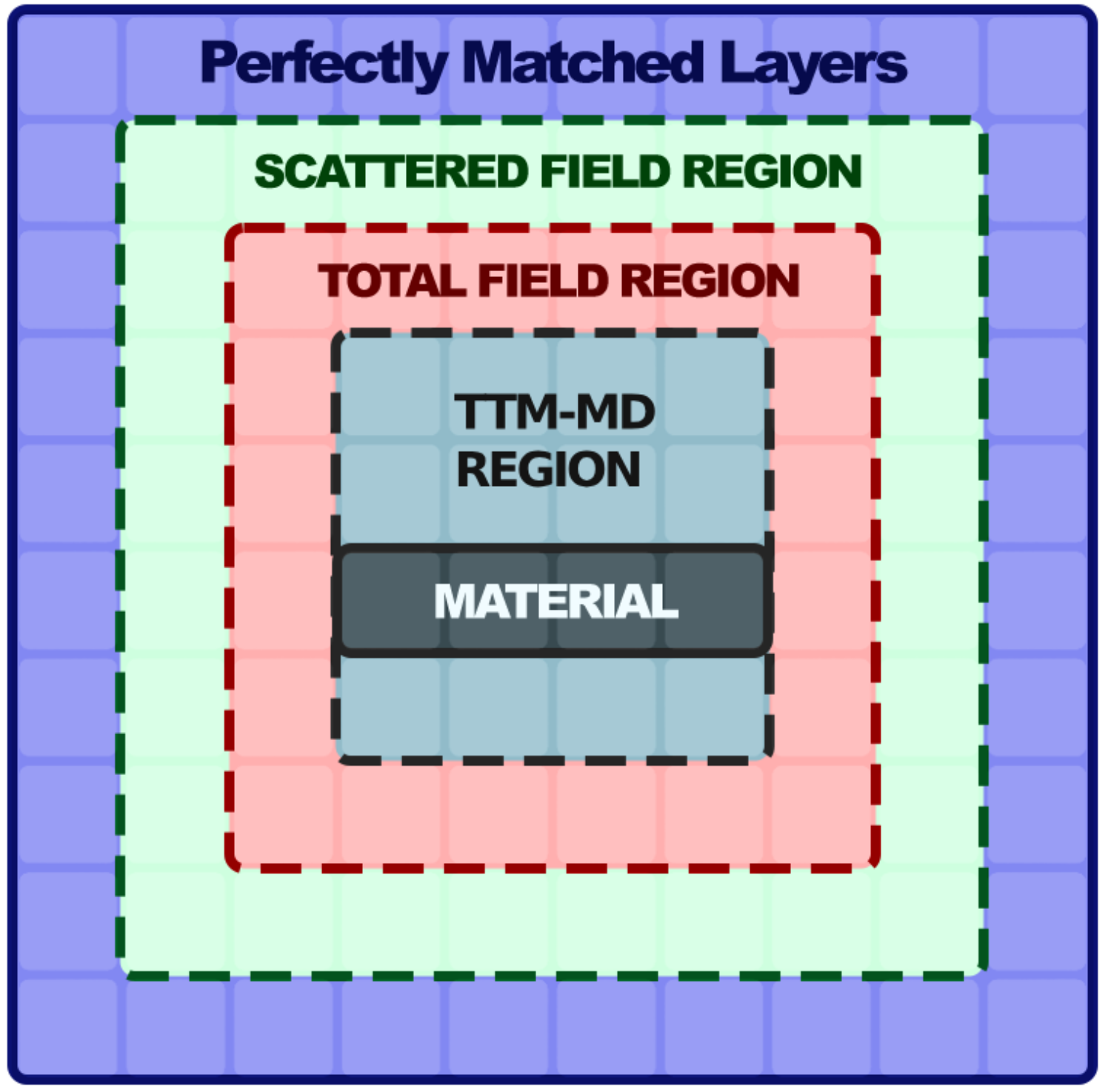}
    \caption{Schematic representation of the computational domain used in the M-TTM-MD model. The domain consists of several regions: the central TTM-MD region represents the simulated material. The surrounding total field region (red dashed border) contains the incident electromagnetic field interacting with the material, while the scattered field region (green dashed border) accounts for the outgoing scattered waves. The outermost Perfectly Matched Layers (blue shaded region) absorb outgoing waves to prevent reflections back into the simulation box.}
    \label{fig:M-TTM-MD-grid-pmls}
\end{figure}

The computational domain{ illustrated in Fig.~\ref{fig:M-TTM-MD-grid-pmls}} comprises several regions: the central TTM-MD zone containing the material, surrounded by the total field (TF) region and the scattered field (SF) zone, both enclosed by the CPMLs when using absorbing boundaries. Periodic boundary conditions can also be used for the lateral boundaries in the FDTD solver for certain configurations, and are mostly retained in the TTM-MD module to emulate an infinite lattice. This framework ensures a self-consistent simulation of ultrafast energy absorption, transport, and structural response within the M-TTM-MD model.

\subsection{Calculating the energy absorption}

Once the FDTD and TTM-MD modules are initialized with the aforementioned considerations, the M-TTM-MD algorithm enters its main loop, where all quantities—temperature distributions, interatomic forces, particle positions and velocities, and electromagnetic field components—are iteratively updated in a self-consistent manner (see Fig.\ref{fig:MTTMMD-interactions}). In this coupled cycle, the
electromagnetic field drives electron heating through the  source term
$Q_{\text{las}}$  {. The energy carried by the field is described by the
Poynting vector $\vec{S}=\vec{E}\times\vec{H}$.
}

{The negative divergence of } $\vec{S}$   {gives the net electromagnetic power entering a local volume, per unit volume. During a transient calculation, this power can both increase the energy stored in the field and material polarization and supply irreversible heating. These contributions must therefore be distinguished in a dispersive medium
\mbox{
\cite{nunes2011electromagnetic}}\hskip0pt
. Within the material, away from imposed
electromagnetic sources and absorbing boundaries, the energy balance for
fixed material coefficients is
}\begin{equation}{\label{eq:FDTD-source-term}
    -\vec{\nabla}\cdot\vec{S}
    = \frac{\partial}{\partial t}
      \left(u_{\rm EM}+u_{\rm ADE}\right)+q_{\rm irr},
    \qquad Q_{\text{las}}=S(\vec{r},t)=q_{\rm irr}.
}\end{equation} 
  {Here, $u_{\rm EM}$ is the electromagnetic energy density, $u_{\rm ADE}$
is the energy density stored in the Drude-Lorentz response, and
$q_{\rm irr}$ is the irreversible dissipated power per unit volume.
In the present coupling, this dissipated power is supplied to the
electron-energy equation, consistent with the two-temperature treatment
of laser heating \mbox{
\cite{anisimov1974electron}}\hskip0pt
. Energy temporarily stored
in the polarization is retained in the optical calculation. It can be
returned to the electromagnetic field or subsequently converted into
heat through damping.
}

{The Poynting flux is evaluated } on the discrete Yee grid   {to describe
energy transport and monitor the optical energy balance. At } cell indices
$(i,j,k)$ and   {time level } $n$,   {its face-centered components are written as
} \begin{equation}
    \pvec{S} =
    \begin{pmatrix}
    \left. S_x \right|^{n}_{i+\frac{1}{2},j,k} \\
    \left. S_y \right|^{n}_{i,j+\frac{1}{2},k} \\
    \left. S_z \right|^{n}_{i,j,k+\frac{1}{2}}
    \end{pmatrix},
\end{equation}
  {The corresponding finite-difference approximation to the net incoming power density, denoted by $Q_{\rm flux}$, is
}\begin{equation}{\label{eq:fdtd-flux-divergence}
    \begin{aligned}
    \left. Q_{\rm flux} \right|^{n}_{i,j,k}
    &= -\frac{\left. S_x \right|^{n}_{i+\frac{1}{2},j,k}
             -\left. S_x \right|^{n}_{i-\frac{1}{2},j,k}}{\Delta x}
       -\frac{\left. S_y \right|^{n}_{i,j+\frac{1}{2},k}
             -\left. S_y \right|^{n}_{i,j-\frac{1}{2},k}}{\Delta y} \\
    &\quad
       -\frac{\left. S_z \right|^{n}_{i,j,k+\frac{1}{2}}
             -\left. S_z \right|^{n}_{i,j,k-\frac{1}{2}}}{\Delta z}.
    \end{aligned}
}\end{equation} 
 {Unlike $Q_{\text{las}}$, this quantity includes changes in stored optical energy and is therefore not used directly as the instantaneous heating source.
} 

Each component of $\vec{S}$ is derived from interpolated electric and magnetic field values.   {Because the Yee scheme staggers these fields in both space and time, the components used in each product must first be evaluated at a common position and time \mbox{
\cite{yee1966numerical}}\hskip0pt
.
For example,
} \begin{align}
\left. S_x \right|^{n}_{i+\frac{1}{2},j,k} = \left. \breve{E}_y \right|^{n}_{i+\frac{1}{2},j,k} \left. \breve{H}_z \right|^{n}_{i+\frac{1}{2},j,k}
- \left. \breve{E}_z \right|^{n}_{i+\frac{1}{2},j,k} \left. \breve{H}_y \right|^{n}_{i+\frac{1}{2},j,k},
\end{align}
  {The spatial interpolation is obtained by averaging } adjacent Yee-grid
  {values. For the electric-field components in this expression,
} \begin{align}
\left. \breve{E}_y \right|^{n}_{i+\frac{1}{2},j,k} &= \frac{1}{4}\left(\left. E_y \right|^{n}_{i+1,j+\frac{1}{2},k} + \left. E_y \right|^{n}_{i,j+\frac{1}{2},k} + \left. E_y \right|^{n}_{i+1,j-\frac{1}{2},k} + \left. E_y \right|^{n}_{i,j-\frac{1}{2},k}\right), \nonumber\ \\
\left. \breve{E}_z \right|^{n}_{i+\frac{1}{2},j,k} &= \frac{1}{4}\left(\left. E_z \right|^{n}_{i+1,j,k+\frac{1}{2}} + \left. E_z \right|^{n}_{i,j,k+\frac{1}{2}} + \left. E_z \right|^{n}_{i+1,j,k-\frac{1}{2}} + \left. E_z \right|^{n}_{i,j,k-\frac{1}{2}}\right). \nonumber
\end{align}
The  {magnetic-field components are likewise interpolated in space and
averaged between their neighboring half-time levels to obtain values
at $t_n$. The } remaining components
$\left. S_y \right|^{n}_{i,j+\frac{1}{2},k}$ and
$\left. S_z \right|^{n}_{i,j,k+\frac{1}{2}}$ are computed analogously.

  {The heating source is calculated from the irreversible damping losses of
the ADE polarization update. These losses are accumulated over the FDTD
substeps before transfer to } the TTM grid  {. This avoids treating reversible
polarization work as heat, which would also occur if the instantaneous
quantity $\vec{E}\cdot\vec{J}$ were used without separating its storage
contribution. The discrete energy balance uses the storage and loss
expressions associated with the implemented ADE recurrence. If material
coefficients change, the associated work is accounted for separately,
as discussed in Sec.~\ref{sec:energy-balance}.
}

{Energy transfer between the FDTD and TTM grids is based on their cell-overlap volumes and, as mentioned in the previous section, does not require an integer ratio between their spacings. }   {For the special case of aligned, uniform cubic grids with }  $\Delta_{TTM}=m\Delta_{FDTD}$,  {where } $m\in\mathbb{N}^{*}$,  {each TTM cell contains $m^3$ complete FDTD cells, and the general overlap-weighted transfer reduces to simple volume averaging. The deposited power density }   {is therefore volume-averaged} over $m^3$ cells before being transferred to the corresponding TTM cell   {. For equal-volume FDTD cells, this gives}\begin{equation}
    {\overline{q}_a=\frac{1}{m^3}\sum_{f\subset a}q_f,
    \qquad \Delta E_a=V_a\overline{q}_a\Delta t,
}\end{equation}
{where $q_f$ is the irreversible power density averaged over the transfer interval $\Delta t$, and $V_a$ is the TTM-cell volume} . This averaging   {preserves the deposited energy while allowing FDTD to resolve finer optical features than the thermal grid.}

{For grids that do not have this simple alignment, the transfer is expressed using the overlap volume $V_{af}$ between FDTD cell $f$ and thermal cell $a$:
}\begin{equation}{\label{eq:source-map}
    \Delta E_a=\sum_f V_{af}q_f\Delta t,
    \qquad
    \sum_a\Delta E_a=\sum_f V_fq_f\Delta t.
}\end{equation}
{The second equality holds when the receiving cells cover the source volume without gaps or double counting. Changes in source coverage caused by material motion, together with any redistributed or exported energy, are recorded separately in the energy balance.
}

 \subsection{The TTM-MD Scheme}

Once the external field's source term $S(\vec{r},t)$ is calculated, the TTM-MD module takes over the calculations. In this scheme, the energy balance is achieved by linking the TTM diffusion equation for electrons to the MD equations of motion \cite{ivanov2003combined}:
\begin{subequations}\label{eq:TTM-MD}
  \begin{align}
{C_e(T_e)\partial_t T_e }&{= \nabla\cdot\left(K_e(T_e,T_l)\nabla T_e\right) - G(T_e - T_l) + S(\vec{r},t), \label{eq:TTM}}\\
{m_i\frac{\partial^2 \vec{r}_i}{\partial t^2} }&{= \vec{F}_i + \xi m_i \vec{v}_i^T, \label{eq:MD}
}\end{align} 
\end{subequations}
where the coupling factor $\xi$ is defined as:
\begin{equation}
\xi = \frac{1}{n}\sum_{k=1}^{n}\frac{G V_N (T_e^k - T_l)}{\sum_i m_i (\vec{v}_i^T)^2}.
\end{equation}
The coupling term distinguishes atomic thermal motion $\vec{v}_i^T$ from the collective motion, allowing temperature-dependent energy transfer without explicit lattice heat capacity. The domain is divided into finite-difference cells where the energy exchanged over $n$ steps is $\Delta E^{\text{e-ph}} = \sum_{k=1}^{n}\Delta t_{\text{FD}} G V_N (T_e^k - T_l)$, which is added to the atomic subsystem through Eq.~\eqref{eq:MD}.

{For gold, the temperature-dependent electron heat capacity $C_e(T_e)$ and electron--phonon coupling factor $G(T_e)$ are taken from Lin et al.~\cite{lin2008electron}, who evaluated these properties using the electronic density of states obtained from density functional theory (DFT) calculations. The electron thermal conductivity $K_e(T_e,T_l)$ follows the model of Anisimov and Rethfeld~\cite{anisimov1997theory}.}

\begin{figure}[h!]
\centering
\includegraphics[width=1\linewidth]{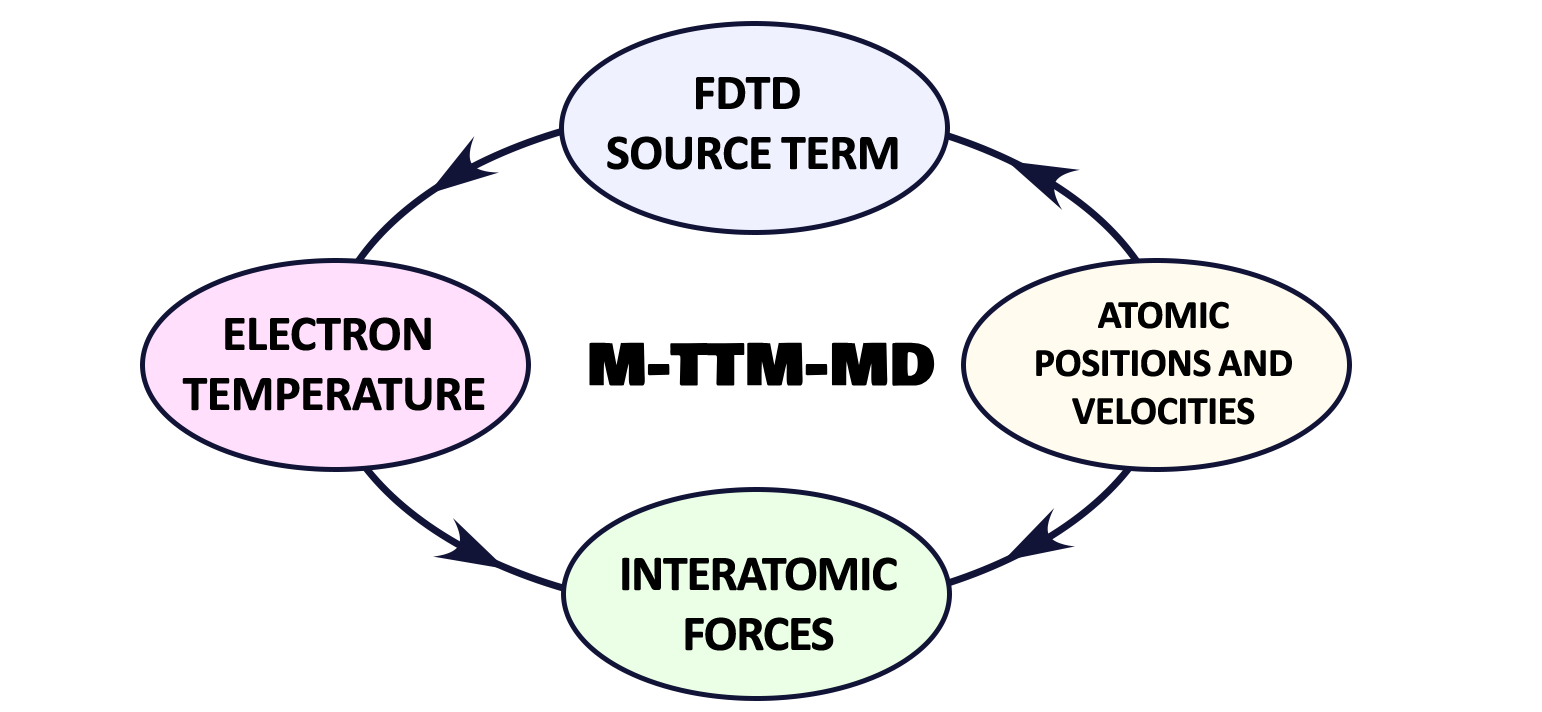}
\caption{{Schematic illustration of the M-TTM-MD cycle, highlighting the interaction between the different subroutines. The FDTD source term drives the electron temperature elevation, influencing the lattice temperature via electron--phonon (e-ph) coupling. The increase in lattice temperature causes matter reorganization, altering the interatomic forces acting on individual particles. These forces are used to compute new positions and velocities of the particles, leading to updated atomic configurations. If the surface topology of the material changes, the scattering of the electromagnetic (EM) pulse is also affected, leading to further updates in the energy deposition profile and completing the cycle.}}

\label{fig:mmdttm_scheme}
\end{figure}

Lattice thermophysical properties—heat capacity, elastic constants, expansion, and melting temperature—are obtained from Embedded-Atom Method (EAM) potentials \cite{zhukhovitskii2020thermodynamics}. The kinetic temperature of atoms (which is the same as the thermodynamic temperature, in equilibrium) can be found using the equipartition theorem:
  \begin{equation}
{T_{\text{kin}} = \frac{2}{k_B(3N - N_c)} \frac{1}{2}\sum_{i=1}^{N} m_i |\vec v_i^T|^2,
}\end{equation} 
where  {the thermal velocity subtracts the local collective velocity, } $N_c$   {counts the removed or constrained degrees of freedom, } and $N$ is the  number of atoms   {in the averaging cell. The electron-to-atom energy transfer is additionally checked against the work of the discrete coupling force.
The } pressure on the other hand is computed via the virial theorem:
\begin{equation}
P = \frac{N k_B T}{V} + \frac{1}{3V} \left\langle \sum_{i=1}^{N} \vec{r}_i \cdot \vec{F}_i^{\text{int}} \right\rangle. 
\end{equation}
 {This formulation allows the TTM-MD model to self-consistently track electron--phonon coupling, lattice response, and phase transitions during ultrafast laser excitation \cite{ivanov2003combined,ivanov2019numerical} as is represented schematically in Figure~\ref{fig:mmdttm_scheme}.}

\subsection{Algorithm of the numerical simulation}

As mentioned previously, the M-TTM-MD algorithm integrates three main numerical methods: FDTD, TTM and MD simulations. To execute at scale, we employ MPI-based domain decomposition that distributes the FDTD, TTM, and MD workloads across many processes (ranks). Each rank advances its local spatial subdomain and exchanges  {the boundary data required by neighboring subdomains during the updates of its solver.} The FDTD and TTM-MD modules  {can use different MPI decompositions. The deposited energy is transferred from the FDTD ranks to the ranks that own the receiving TTM cells. This exchange uses the conservative mapping described above, together with the ownership information that identifies the receiving rank for each thermal cell.} 
\par In practice, we implement \emph{{halo}} layers{} that extend each rank's subdomain to support  {calculations near process boundaries. These layers store copies of data owned by neighboring ranks. Their extent depends on the field stencil or atomic interaction range used by the corresponding solver.} In the FDTD mesh, the {halo}{} layer stores $\vec{E}$ and $\vec{H}$ components one cell past each interface and is refreshed from neighboring ranks at every {FDTD } timestep, ensuring consistent curl updates at rank interfaces  \mbox{\cite{su2004novel}}. 
\par In the {MD module}, an analogous {ghost-atom region}{} caches atoms from adjacent ranks within the neighbor cutoff; this buffer supports force evaluation near subdomain edges . Atoms that cross a boundary are handed off to the appropriate rank at the next communication step. {The TTM grid likewise exchanges the neighboring thermal values needed to evaluate heat diffusion across subdomain boundaries.} This design {distributes the computational work}, reduces the memory footprint per rank, and  {allows the coupled solvers to advance their assigned subdomains in parallel.}
\par{Each solver advances with its own time step. Energy and material data are exchanged at synchronization points that correspond to the same physical time in the coupled calculation. Several FDTD or thermal substeps can therefore occur within one coupling interval.} 
\par
\begin{figure}[h!]
\centering
\includegraphics[width=1\linewidth]{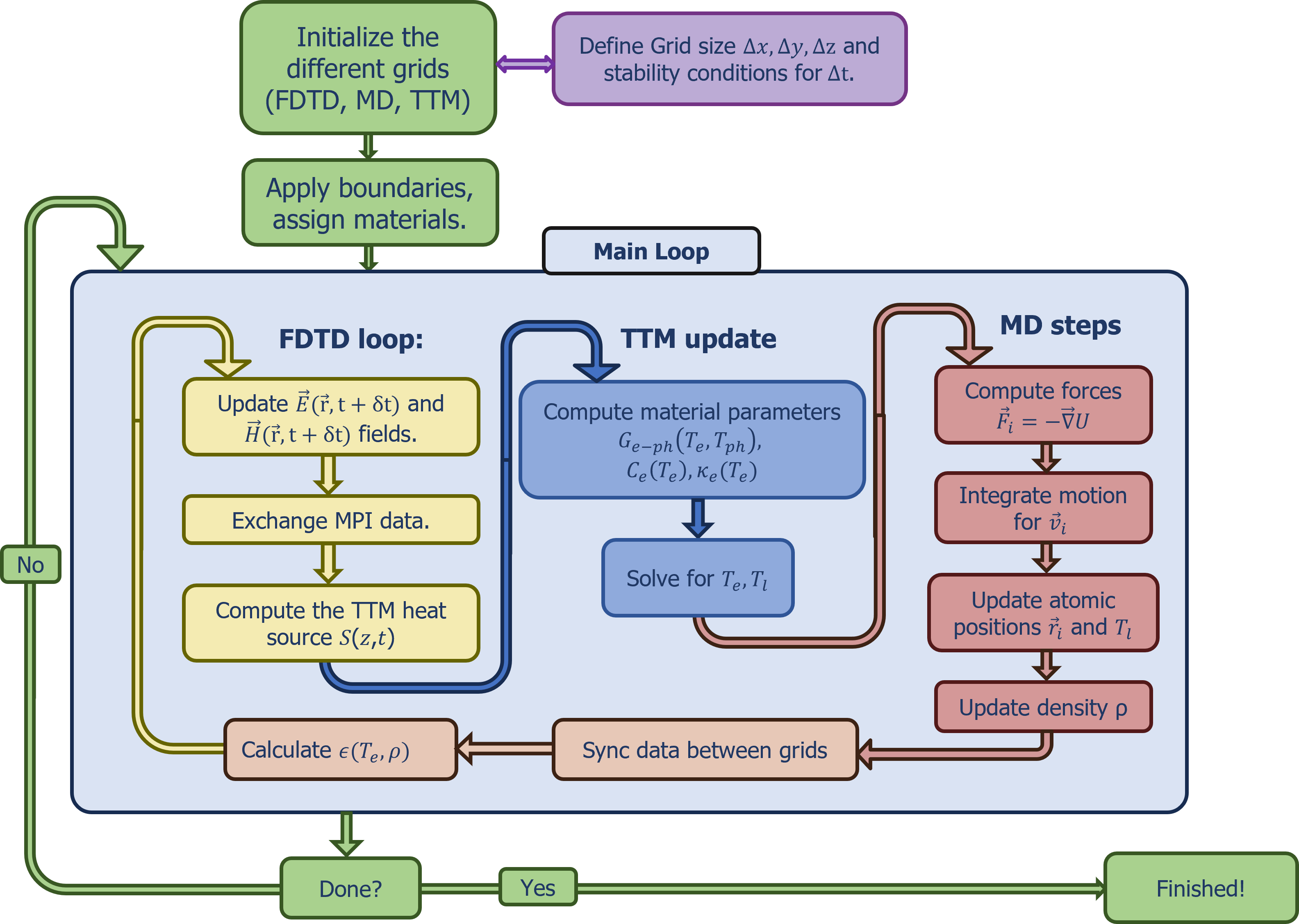}
\caption{Basic implementation of the M-TTM-MD algorithm. The FDTD module, the TTM energy equation, and the MD integrator exchange data at each global timestep, ensuring a self-consistent coupling between the different components.}
\label{fig:M-TTM-MD-scheme-impl}
\end{figure}

 The general outline of the combined algorithm{, shown in Fig.~\ref{fig:M-TTM-MD-scheme-impl},} can be described as follows: 
\begin{enumerate} 
    \item Initialize the MPI libraries and set parameters for parallelization. 
    \item Read the values from input files, including those for simulation parameters and field initialization. {Load the material properties and initial atomic configuration.} 
    \item Set up the {MD neighbor-cell structure and the TTM and FDTD meshes}{} and apply  {the prescribed boundary conditions for each module, including periodic, open, or absorbing boundaries where specified.} 
    \item Broadcast relevant variables to all processors. 
    \item Initialize the FDTD method and  {assign the optical material fractions from the atomic geometry. Construct the mapping used to transfer deposited energy to the TTM cells. For a restarted calculation, restore the saved physical state and time-step counters of the active modules.} 
    \item Start the main time loop over $t_{\text{global}}$: 
    \begin{enumerate} 
        \item {While the optical integration window is active, } run the FDTD loop until  {the fields reach the next coupling synchronization time:} 
        \begin{enumerate} 
            \item Calculate the electromagnetic source field.
             \item Update the FDTD equations by calculating the new electromagnetic fields {and advancing the ADE polarization variables}. 
             \item Exchange data between processors as required. 
             \item Calculate the electromagnetic energy flux. {Accumulate the irreversible ADE dissipation and record the electromagnetic and polarization energy needed for the optical energy balance.} 
             \item Write FDTD variables to external files for output and potential resumption {at the prescribed output and checkpoint intervals}. 
        \end{enumerate} 
        \item {Transfer the accumulated deposited energy to the TTM cells. Advance electron heat diffusion and electron--phonon energy exchange using the required thermal substeps.} 
        \item Update the MD algorithm by recalculating the new forces, velocities, and atomic positions. {Include the energy exchange with the electron subsystem through the TTM-MD coupling.} 
        \item  {Evaluate the lattice temperature from the updated atomic velocities and update the material geometry. Refresh the optical material description and transfer map at the prescribed material-update points. Record the energy carried by any atoms leaving through an open boundary.} 
        \item Write relevant variables to external files. {Record the coupled energy balance and save restart data at the prescribed intervals.} 
    \end{enumerate} 
    \item Once the main time loop is complete, finalize the program and close all processes. 
\end{enumerate} {After the optical integration window ends, the TTM-MD calculation continues to the prescribed final time. This stage describes the subsequent heat transport, electron--phonon relaxation, and atomic motion.}
\subsection{Energy balance}
\label{sec:energy-balance}

Beyond numerical consistency, detailed energy accounting reveals how energy redistributes across subsystems during nonequilibrium processes (e.g., laser-induced melting). Accordingly, conservation is verified for each algorithmic component. {The checks cover the optical solver, energy transfer between meshes, and the coupled TTM-MD subsystem.}

For the FDTD subroutine with a TFSF plane-wave source, we integrate the incident power and compare {the resulting incident energy with the reflected, transmitted, and absorbed energies, including any change in the energy stored in the fields and material polarization}. The incident {energy is determined from a matched vacuum calculation using the same source and flux-monitor configuration}; the scattered-field region enables {the reflected energy to be evaluated, while the total transmitted flux is measured beyond the sample}, and the material volume integral {of the irreversible dissipation } yields the absorbed part. {Reflection and transmission include all propagating diffraction orders. For lateral periodic boundaries, the fluxes through opposite lateral faces cancel.}

An alternative check {evaluates} the electromagnetic energy-balance law {using a collocated finite-difference diagnostic } on the Yee grid. While Yee staggering yields accurate FDTD updates, evaluating $\vec{S}$, $\vec{E}\cdot\vec{J}$, and {the electromagnetic energy density $u_{\rm EM}$ at a common location and time} requires {spatial and temporal} interpolation. {For fixed material coefficients, the electromagnetic energy density is}
\begin{equation}
{u_{\rm EM}=\frac{\epsilon_0\epsilon_\infty}{2}|\vec{E}|^2+\frac{\mu_0}{2}|\vec{H}|^2.}
\end{equation}
{Here $\epsilon_\infty$ is the real background relative permittivity. Energy stored in the dispersive polarization is calculated separately from the ADE variables.}

The {local field balance and its collocated diagnostic} read:
\begin{align}
&\frac{\partial {u_{\rm EM}}}{\partial t} + \vec{\nabla}\cdot \vec{S} = - \vec{E} \cdot \vec{J} \nonumber \\
&\sum_{x_i} \epsilon_0 \epsilon_\infty \frac{\left. \breve{E}^2_{x_i} \right|^{n+\frac{1}{2}}_{i,j,k} - \left. \breve{E}^2_{x_i} \right|^{n-\frac{1}{2}}_{i,j,k}}{2 \Delta t} \nonumber \\
&\quad + \sum_{x_i} {\mu_0} \frac{\left. \breve{H}^2_{x_i} \right|^{n+\frac{1}{2}}_{i,j,k} - \left. \breve{H}^2_{x_i} \right|^{n-\frac{1}{2}}_{i,j,k}}{2  \Delta t} \nonumber \\
&\quad + \frac{\left. S_{x} \right|^{n}_{i+\frac{1}{2},j,k} - \left. S_{x} \right|^{n}_{i-\frac{1}{2},j,k}}{\Delta x} + \frac{\left. S_{y} \right|^{n}_{i,j+\frac{1}{2},k} - \left. S_{y} \right|^{n}_{i,j-\frac{1}{2},k}}{\Delta y} \nonumber \\
&\quad + \frac{\left. S_{z} \right|^{n}_{i,j,k+\frac{1}{2}} - \left. S_{z} \right|^{n}_{i,j,k-\frac{1}{2}}}{\Delta z} \nonumber \\
&= - \sum_{x_i} \left. \breve{E}_{x_i} \right|^{n}_{i,j,k} \left. \breve{J}_{x_i} \right|^{n}_{i,j,k}{+\left.r_{\rm coll}\right|^n_{i,j,k}},
\end{align}
with $x_i \in \{x,y,z\}$ and
\begin{equation}
\left. \breve{J}_{x_i} \right|^{n}_{i,j,k} = \left. \breve{J}_{D,x_i} \right|^{n}_{i,j,k} + \sum_{l=1}^{L_p} \left. \breve{J}_{l,x_i} \right|^{n}_{i,j,k},
\end{equation}
where $\breve{\cdot}$ denotes interpolated quantities; e.g.,
\begin{align*}
\left. \breve{E}_{x} \right|^{n+\frac{1}{2}}_{i,j,k} &= \frac{1}{4} \left(\left. E_{x} \right|^{n+1}_{i+\frac{1}{2},j,k} + \left. E_{x} \right|^{n}_{i+\frac{1}{2},j,k} + \left. E_{x} \right|^{n+1}_{i-\frac{1}{2},j,k} + \left. E_{x} \right|^{n}_{i-\frac{1}{2},j,k}\right), \\
\left. \breve{H}_{x} \right|^{n+\frac{1}{2}}_{i,j,k} &= \frac{1}{4} \left(\left. H_{x} \right|^{n+\frac{1}{2}}_{i,j+\frac{1}{2},k+\frac{1}{2}} + \left. H_{x} \right|^{n+\frac{1}{2}}_{i,j+\frac{1}{2},k-\frac{1}{2}} + \left. H_{x} \right|^{n+\frac{1}{2}}_{i,j-\frac{1}{2},k+\frac{1}{2}} + \left. H_{x} \right|^{n+\frac{1}{2}}_{i,j-\frac{1}{2},k-\frac{1}{2}}\right).
\end{align*}

{The residual $r_{\rm coll}$ measures the imbalance obtained with these interpolated quantities. Its value depends on the spatial and temporal interpolation. The discrete solver energy check uses storage and flux terms consistent with the field and ADE updates. Over a common integration interval, its residual is}
\begin{equation}
\label{eq:optical-ledger}
{r_F=\Delta U_{\rm EM+ADE}+E_{\rm outward}+Q_{\rm ADE}+W_{\rm Ohmic}-W_{\rm coeff}.}
\end{equation}
{Here $\Delta U_{\rm EM+ADE}$ is the change in electromagnetic and reversible polarization energy, $E_{\rm outward}$ is the signed net outward energy flux, and $Q_{\rm ADE}$ is the irreversible energy dissipated through the ADE damping terms. The boundary-flux term includes the source contribution for the chosen control volume. The term $W_{\rm Ohmic}$ accounts for any additional nondispersive conductivity loss, and $W_{\rm coeff}$ is the work associated with changes in the material coefficients, taken as positive when energy is added. 

}

{The corresponding full-pulse balance, expressed in terms of the independently measured incident, reflected, and transmitted energies, is}
\begin{equation}
\label{eq:integral-ledger}
{E_{\rm inc}+W_{\rm coeff}=E_R+E_T+Q_{\rm ADE}+W_{\rm Ohmic}+\Delta U_{\rm EM+ADE}.}
\end{equation}
{The flux monitors lie inside the physical domain, before the CPMLs. Energy leaving through these monitors is counted as outgoing energy; its subsequent absorption occurs outside this control volume. The integrated balance error is normalized by the incident pulse energy $E_{\rm inc}$.}

For the TTM-MD subsystem, the total system energy includes the electron energy $E_e$, {the atomic kinetic energy $K_{\rm MD}$, and the interatomic potential energy $U_{\rm EAM}$. The atomic energies include thermal and collective motion and the energy changes associated with melting} \cite{ivanov2003combined,ivanov2019numerical,zhukhovitskii2020thermodynamics}. {The total matter energy and its residual are}
\begin{equation}
\label{eq:matter-ledger}
{E_M=K_{\rm MD}+U_{\rm EAM}+E_e,\qquad r_M=\Delta E_M+E_{\rm exported}-Q_{\rm TTM}-W_{\rm external}.}
\end{equation}
{The energy residual $r_M$ measures the difference between the change in stored energy, including energy exported from the system, and the total energy supplied over the same interval. It vanishes when all energy contributions are accounted for exactly.
Here $Q_{\rm TTM}$ is the optical energy received by the thermal cells, $W_{\rm external}$ is the prescribed external work, and $E_{\rm exported}$ includes energy carried by removed atoms and any energy exported through thermal boundaries or during remapping. Electron--phonon coupling transfers energy internally between the electron and atomic subsystems. For the FDTD source used here, conservation during source transfer is checked through $r_{\rm map}=Q_{\rm TTM}-Q_{\rm ADE}$. 
}

\section{Numerical Simulations of Thin Au Films}
\label{sec:Au-sim}

\begin{figure}[h!]
\centering
\begin{minipage}[t]{.48\linewidth}
\centering
\includegraphics[width=\linewidth]{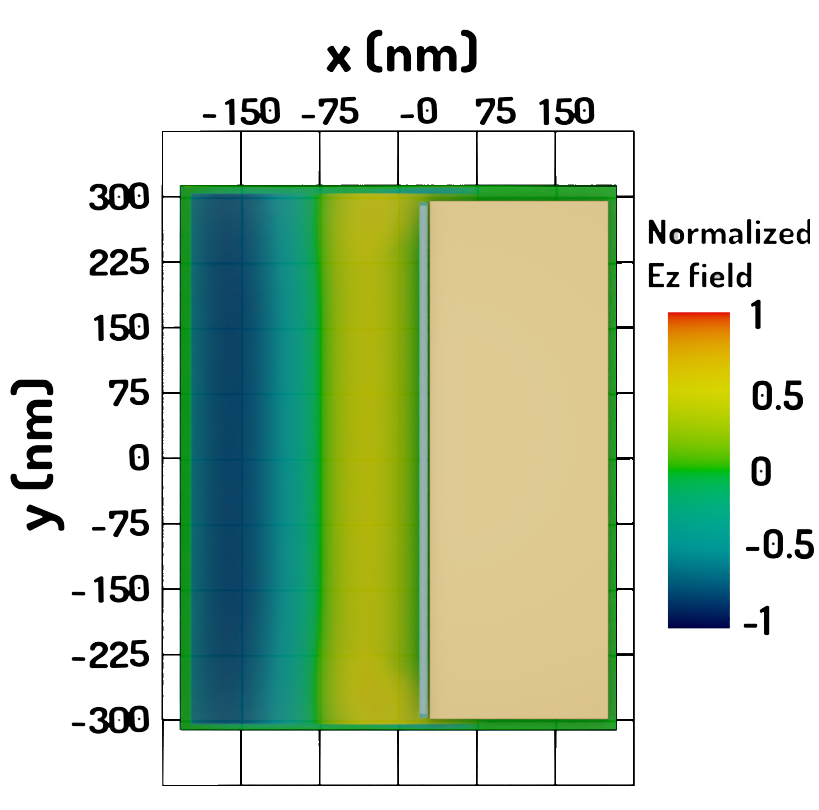}\\
\textbf{(a)}
\end{minipage}
\hfill
\begin{minipage}[t]{.48\linewidth}
\centering
\includegraphics[width=\linewidth]{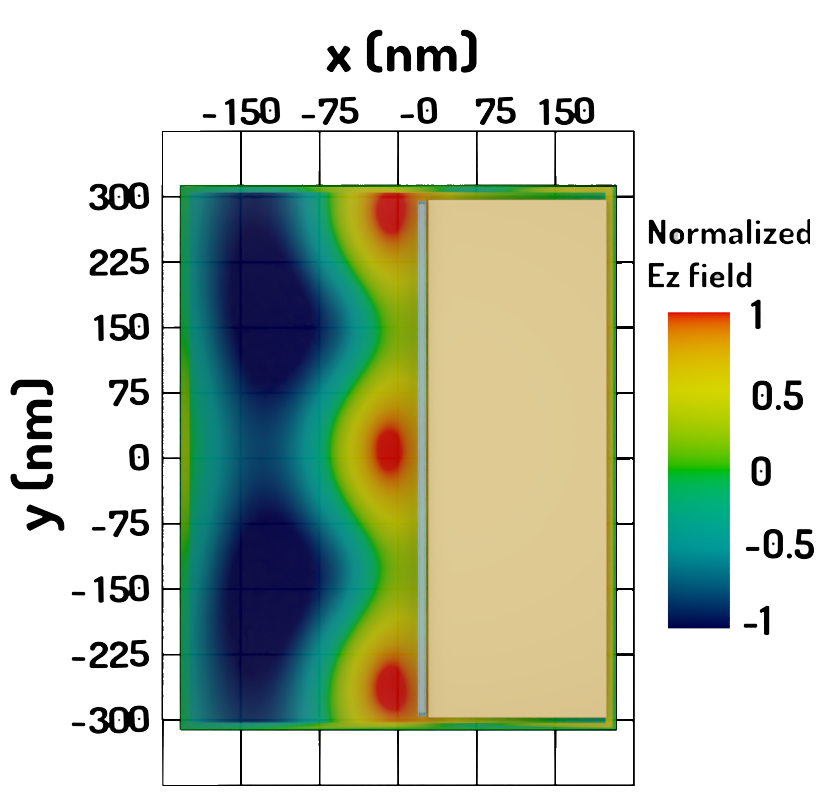}\\
\textbf{(b)}
\end{minipage}
\caption{Normalized electric field $E_z$ during the FDTD run: (a) A normally incident plane wave is injected in the total-field (TF) region and impinges on the Au thin film after being transmitted {through} a thin glass layer {separated from the gold by a vacuum gap}; (b) Interference between incident, reflected, and transmitted waves produces standing-wave fringes that {influence} the spatial absorption profile used by the TTM module.}
\label{fig:M-TTM-MD-Ez-field}
\end{figure}

In this section, we illustrate the M-TTM-MD framework by simulating the irradiation of a $150\times600\times15$~nm Au thin film with six femtosecond laser pulses of 300~nm wavelength, 50~fs duration, and 0.1~J/cm$^2$ fluence, separated by 5~ps intervals to allow partial relaxation between successive excitations. The computational domain (Fig.~\ref{fig:M-TTM-MD-Ez-field}) contains the gold film on the left side and a thin dielectric layer ($\epsilon_r = 3.5$) placed above it with a narrow vacuum gap, {forming a layered dielectric--vacuum--gold geometry}. The FDTD grid resolves both materials with uniform spacing in all spatial directions and using a grid size equal to that utilized in the TTM mesh. The MD algorithm on the other hand contains approximately {$8\times10^7$} atoms in total, simulating only the Au sample and neglecting the glass material.

As shown in Fig.~\ref{fig:M-TTM-MD-Ez-field}(a), a normally incident plane wave is launched into the total-field region of the FDTD domain, producing a standing-wave interference pattern through the superposition of incident, reflected, and transmitted components. This interference gives rise to a spatially modulated electric field distribution at the {metal surface}, which { contributes to } the nonuniform absorption profile transferred to the TTM module. This self-consistent solution of Maxwell's equations enables the algorithm to naturally capture the { geometry-dependent } optical absorption landscape responsible for seeding periodic energy deposition and subsequent structural modification.

\begin{figure}[h!]
\centering
\includegraphics[width=1\linewidth]{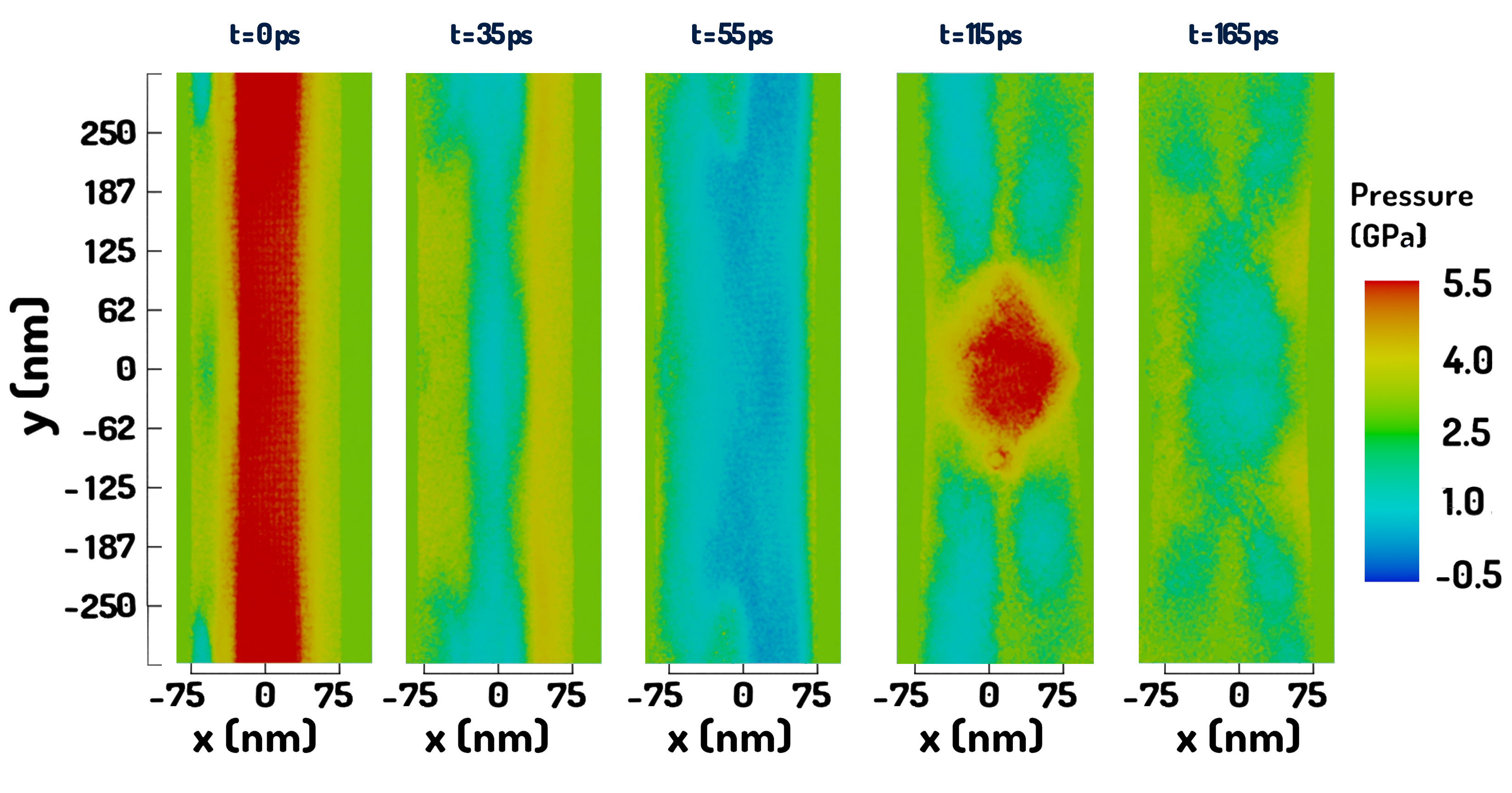}
\caption{Temporal evolution of the lattice pressure slices revealing stress confinement. The evolving stress field tracks the spatio-temporal energy deposition and subsequent response of the thin film.}
\label{fig:M-TTM-MD-Press-Slices}
\end{figure}

During irradiation, the absorbed energy rapidly elevates the electron temperature, while the lattice remains initially cold. The electron temperature profile initially mirrors the optical absorption pattern, exhibiting high temperatures localized at the interference maxima near the surface. As energy diffuses through {electronic heat conduction} within a few hundred femtoseconds, these temperature gradients progressively smooth out by around one picosecond, leading to partial temperature homogenization across the film length. Similarly, the energy transfer to the lattice through electron-phonon coupling starts with the same spatial distribution of the surface electromagnetic fields, and progressively becomes more uniform on a picosecond timescale. Because the pulse duration (50~fs) is much shorter than the material's acoustic relaxation time, the rapid lattice heating leads to a fast build-up of internal stress confined inside the sample. The rapid thermoelastic expansion results in compressive pressures up to 6 GPa and oscillatory expansion waves that originate from the center of the film and propagate towards the free boundaries. {In} our simulations, we used periodic boundary conditions in the lateral directions normal to the incident laser pulse{. The} expansion waves reflect {from the free surfaces } back and interfere at the center of the film as is seen in Fig.~\ref{fig:M-TTM-MD-Press-Slices}. 

\begin{figure}[h!]
\centering
\includegraphics[width=1\linewidth]{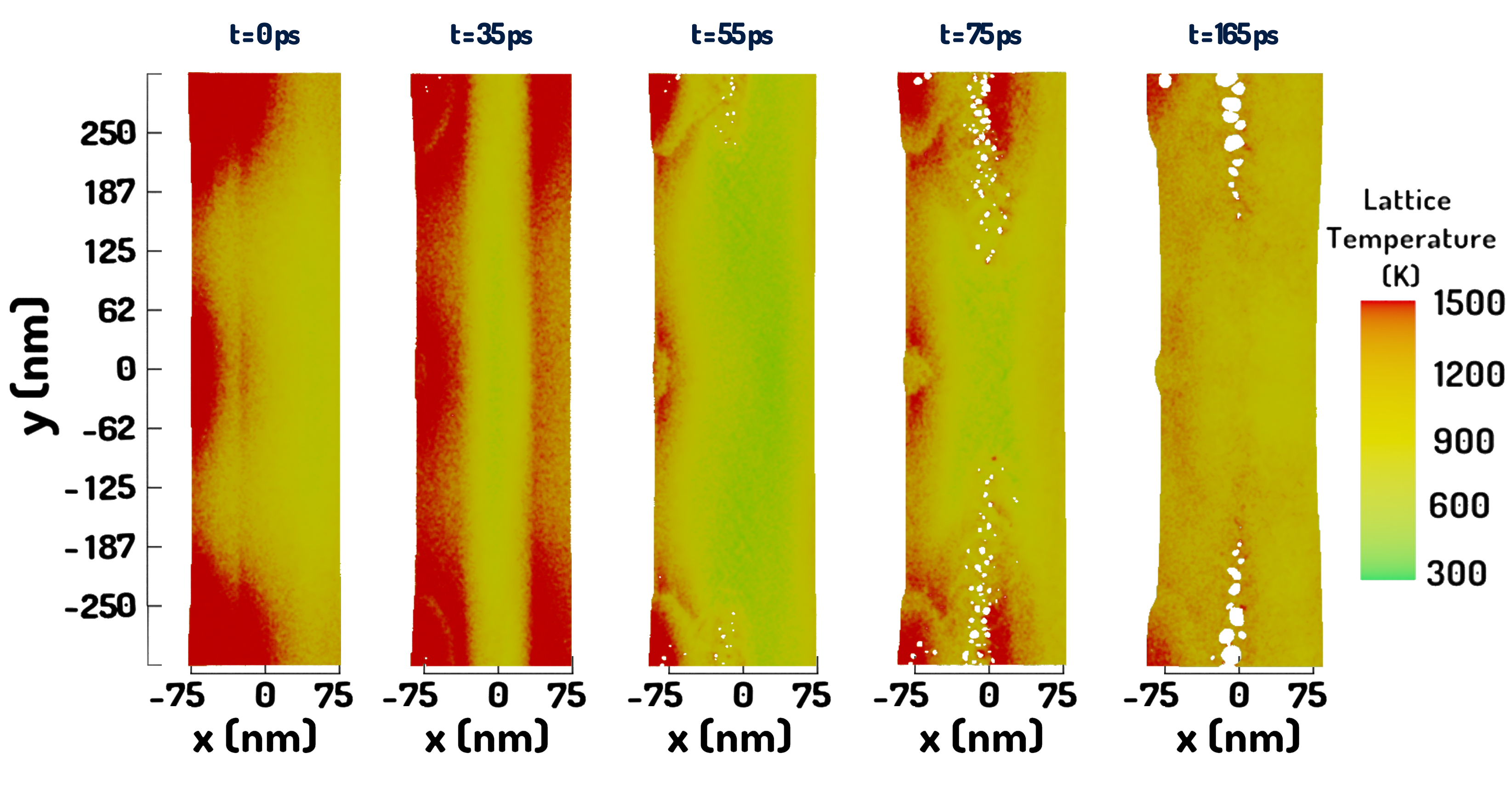}
\caption{Temporal evolution of the lattice temperature $T_\ell$ in the Au film. Heating is initially localized near the {illuminated metal surface}. High pressure regions coincide with elevated temperatures.}
\label{fig:M-TTM-MD-Temp-Slices}
\end{figure}

Following the spatio-temporal evolution of the lattice temperature (Fig.~\ref{fig:M-TTM-MD-Temp-Slices}), we can see that melting initiates heterogeneously at the free surfaces, where the nucleation barrier is lowest, and subsequently propagates toward the film center as heat diffusion and acoustic oscillations continue. In the regions of strongest heating, homogeneous melting {can occur} simultaneously, due to the local superheating the material was exposed to, leading to liquid nucleation and facilitating phase transitions \cite{rethfeld2017modelling,gamaly2011physics,gamaly2013physics}. When the tensile phase of the stress wave traverses the molten regions, the local pressure can drop below the cavitation threshold, causing the spontaneous formation of nanoscopic voids. These voids appear preferentially at the periodic field maxima imposed by the optical interference pattern and represent a mechanical relaxation pathway within the molten layer. The expansion of these voids redistributes the surrounding melt through hydrodynamic flow, pushing material upward around the low-density regions and creating periodic surface elevations (``nanobumps'') aligned with the optical intensity peaks \cite{ivanov2003combined,terekhin2022key}.

\begin{figure}[h!]
\centering
\includegraphics[width=1\linewidth]{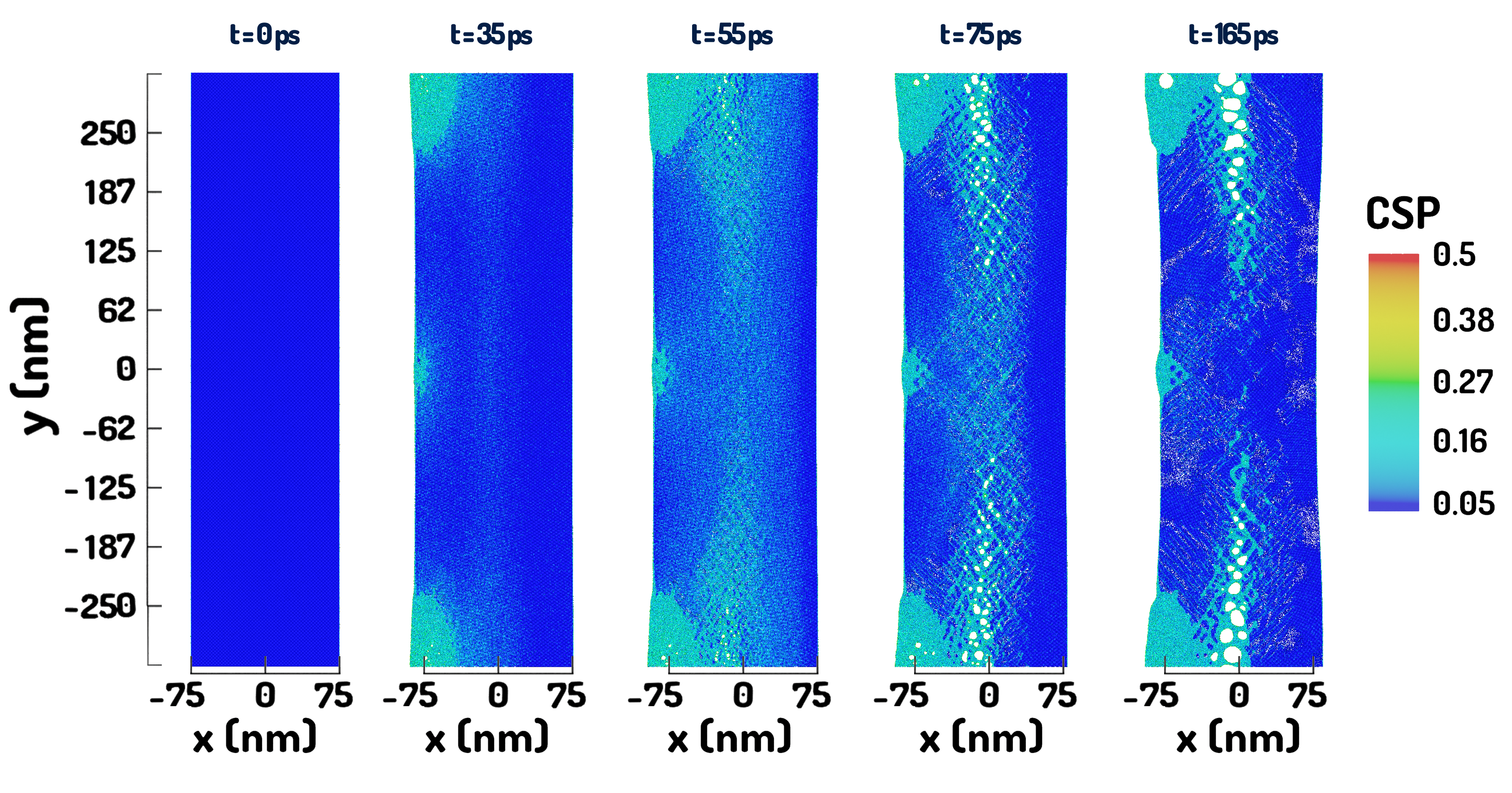}
\caption{Temporal evolution of the irradiated Au film where a thin slice of the full sample is plotted showcasing the atomic structure of the sample. The atoms are colored according to their local {centrosymmetry} parameter, with blue indicating crystalline order and red denoting disordered  states. {Elevated CSP can result from thermal motion, defects, surfaces, or melting; phase identification requires additional structural or thermodynamic information.}}
\label{fig:M-TTM-MD-Fig-Slices}
\end{figure}

As successive pulses arrive, partial thermal relaxation occurs but the overall temperature continues to build up, reinforcing the preexisting modulation pattern. The spatial and temporal evolution of the lattice structure (Fig.~\ref{fig:M-TTM-MD-Fig-Slices}) displays well-defined surface protrusions accompanied by internal voids and dislocation networks that result from the large transient stresses and rapid solid-liquid transitions. The applied fluence lies in the regime where localized melting and cavitation occur without large-scale ablation or sheet detachment{ over the recorded interval}. Overall, the simulation illustrates the sequence of ultrafast processes, from the field-resolved optical interference, nonequilibrium electron-phonon coupling under stress confinement, transient melting and cavitation, to hydrodynamic flow leading to  surface relief---providing a coherent microscopic picture of the initial stages of periodic surface structure formation in thin metallic films.

{

\subsection{Source term comparison between FDTD and Beer--Lambert law}\label{sec:groove-comparison}

In this section, we compare FDTD and Beer--Lambert heating of a grooved gold film to determine how the optical source affects energy deposition and the subsequent structural response. We perform four coupled simulations, using both source models at two incident fluences. In each pair, we use the same relaxed sample and TTM-MD equations, Eq.~\eqref{eq:TTM-MD}, so that we can examine the effect of changing the optical source.

We use a gold film with nominal dimensions $2000\times2.45\times25$ nm, containing approximately 7.1 million atoms. The groove is centered at $x=0$, is 20 nm wide and 12.5 nm deep, and extends along $y$. We apply periodic boundaries in $x$ and $y$ to both the optical and atomistic domains, eliminating the outer lateral edges and giving a groove separation of 2 $\mu$m. The gold surfaces are free along $z$, and CPMLs terminate the optical domain in that direction. The narrow periodic width gives a quasi-two-dimensional representation and restricts transverse deformation.

We irradiate the film at normal incidence with a 550 nm pulse of 40 fs intensity FWHM, using incident fluences of 70 and 100 mJ/cm$^2$. The pulse propagates along $-z$, with the electric field polarized across the groove along $x$. We use a 2.5 nm FDTD grid and follow the TTM-MD evolution to 100 ps with a 2.5 fs MD timestep. The calculations use 384 MPI processes for TTM-MD and 96 for FDTD. The recorded wall times are approximately 4 h for Beer--Lambert and 7--7.5 h for FDTD.

For the field-resolved calculation, we use the heating term defined in Eq.~\eqref{eq:FDTD-source-term}. For normal incidence, we write the Beer--Lambert source as
\begin{equation}
\label{eq:BL-source}
S_{\rm BL}(s,t)=(1-R)\alpha I_{\rm inc}(t)e^{-\alpha s},
\end{equation}
\begin{equation}
\tilde n=\sqrt{\epsilon_{\rm Au}},
\qquad
R=\left|\frac{\tilde n-1}{\tilde n+1}\right|^2,
\qquad
\alpha=\frac{4\pi\,\mathrm{Im}\,\tilde n}{\lambda}.
\end{equation}
Here, $s$ is measured into the material from the illuminated surface, and we use the passive branch of $\tilde n$. For both sources, we evaluate the same bulk dielectric function, Eq.~\eqref{eq:bulk-dielectric}, at fixed optical temperatures of 300 K. The electron and lattice temperatures continue to evolve in the TTM-MD calculation.

We find that FDTD deposits approximately 22.6\% of the incident energy, compared with 13.0\% for Beer--Lambert, at both fluences. The optical field penetrates the 25 nm gold film and is partially reflected at the rear Au--vacuum interface. The returning wave interferes with the forward-propagating field, modifying the electric-field distribution inside the metal and therefore the absorbed energy. FDTD includes these internal reflections and their interference through the electromagnetic solution. The Beer--Lambert source used here accounts for the front-surface reflectivity and single-pass attenuation. Using the same dielectric function, we also find substantially greater absorption in an independent calculation for a flat 25 nm vacuum/Au/vacuum film when coherent internal reflections are included \cite{byrnes2020multilayer}. This supports thin-film interference as an important contribution to the difference between the two source models.

We also obtained good convergence of the total absorbed energy across different meshes in separate optical tests of the earlier 50 nm thick grooved film.

In Figs.~\ref{fig:groove70} and \ref{fig:groove100}, we compare the deposited-energy profiles along $x$ with atomic slices at 100 ps. At 100 mJ/cm$^2$, we observe greater expansion and internal void formation in the FDTD-driven sample, consistent with stronger heating followed by mechanical relaxation. Since both the total absorbed energy and its spatial distribution change, these results show their combined influence on the atomic response.

\begin{figure}[p]

\centering
\includegraphics[width=\linewidth]{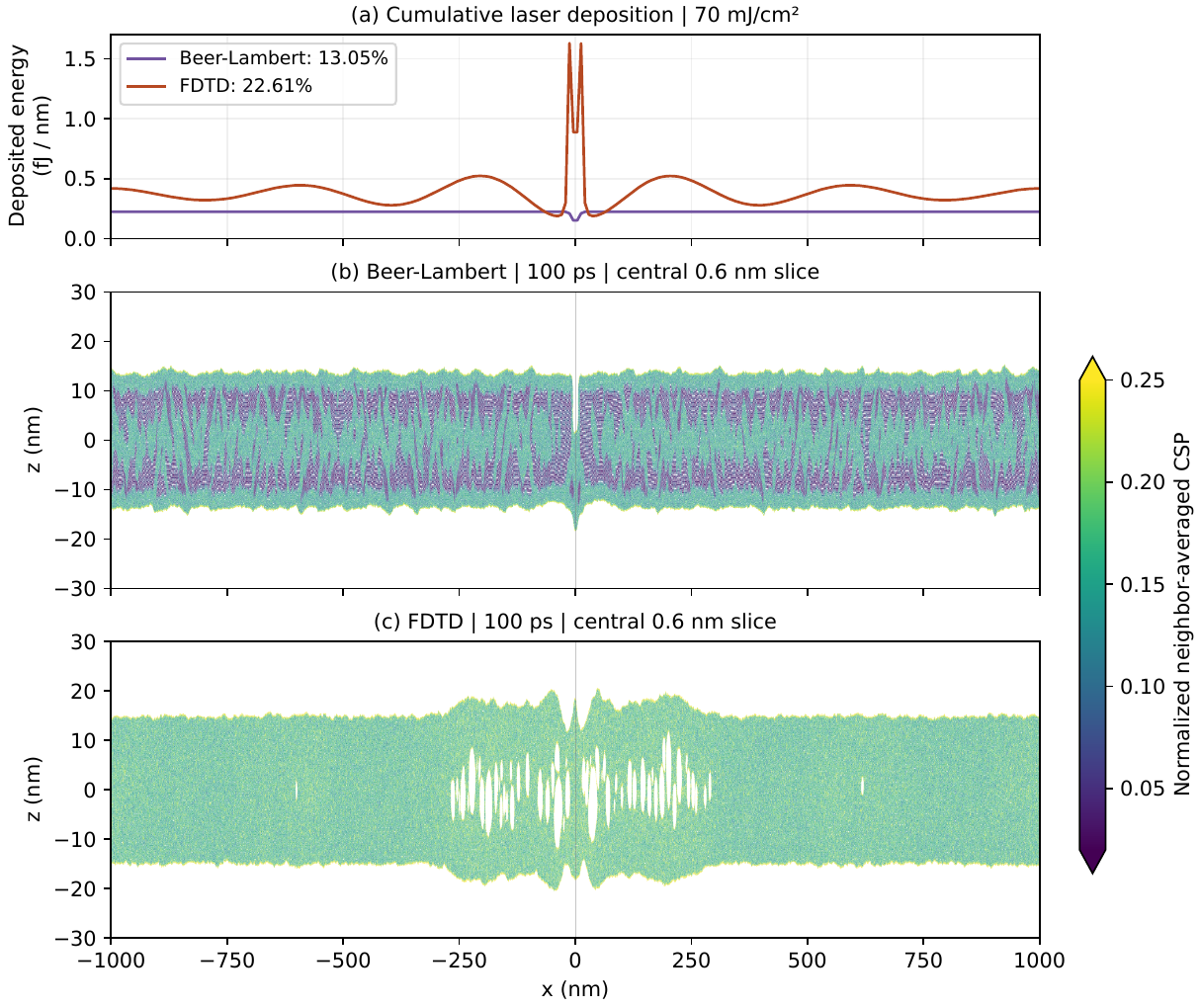}
\caption{Energy deposition and atomic response at 70 mJ/cm$^2$. Top: cumulative deposited energy integrated over $y$ and $z$, and divided by the 8.33 nm bin width along $x$. Lower panels: atomic slices at 100 ps containing all saved atoms with $|y|\leq0.3$ nm. Atoms are colored by the recorded normalized, neighbor-averaged CSP. Both source models use identical spatial and color scales.}
\label{fig:groove70}
\end{figure}

\begin{figure}[p]

\centering
\includegraphics[width=\linewidth]{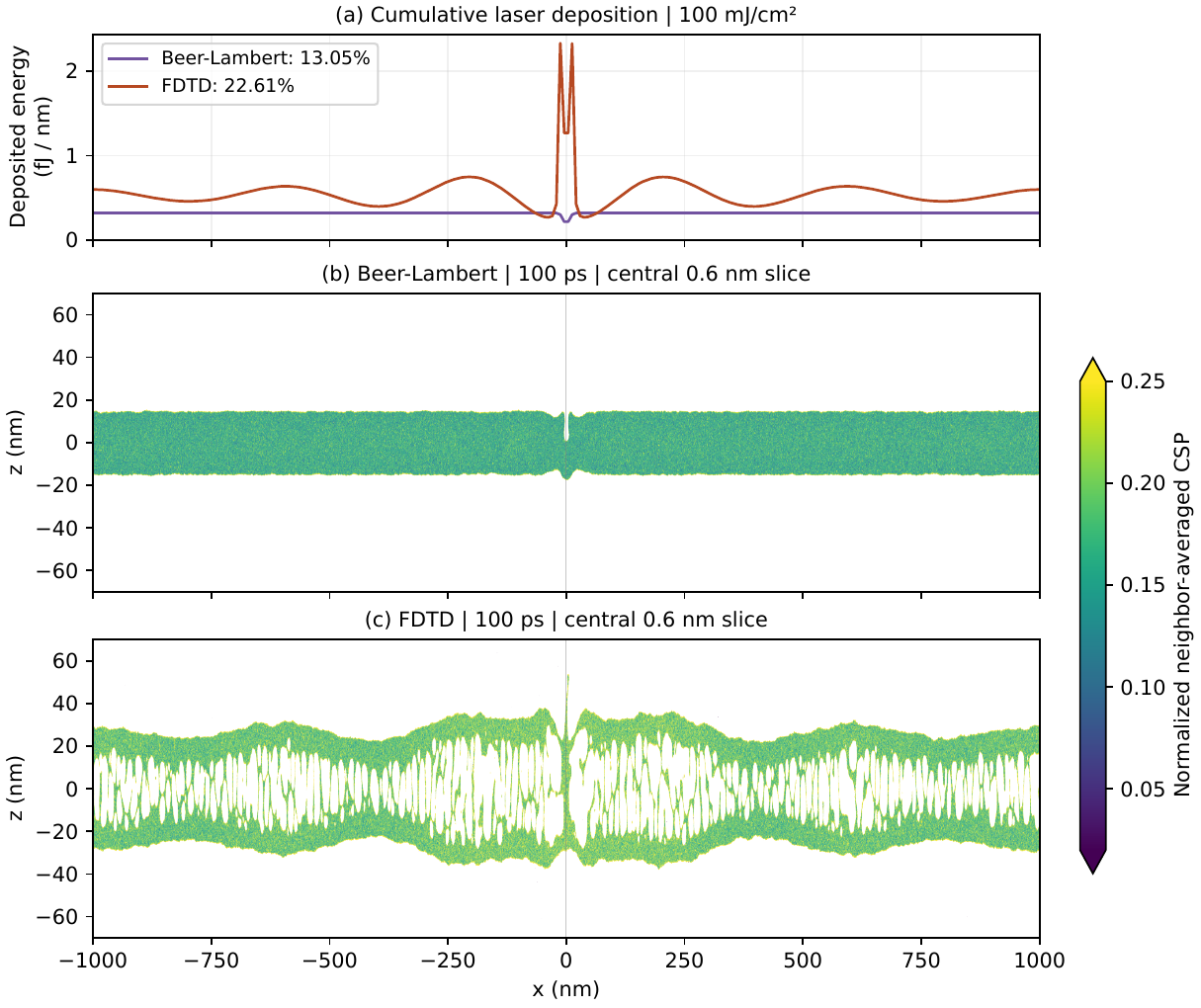}
\caption{Energy deposition and atomic response at 100 mJ/cm$^2$, using the same source-profile and CSP definitions as Fig.~\ref{fig:groove70}. The larger vertical range includes the expanded sample. The FDTD-driven trajectory shows greater expansion and internal void formation than the Beer--Lambert trajectory.}
\label{fig:groove100}
\end{figure}

For the 100 mJ/cm$^2$ FDTD calculation, we show the near-surface electric-field distribution in Fig.~\ref{fig:groove-fields} and the electron and lattice temperature evolution in Fig.~\ref{fig:groove-temperatures}. Together with the deposited-energy profiles and atomic slices, these figures connect the optical response of the groove with the subsequent thermal and structural evolution.

\begin{figure}[htbp]

\centering
\includegraphics[width=\linewidth]{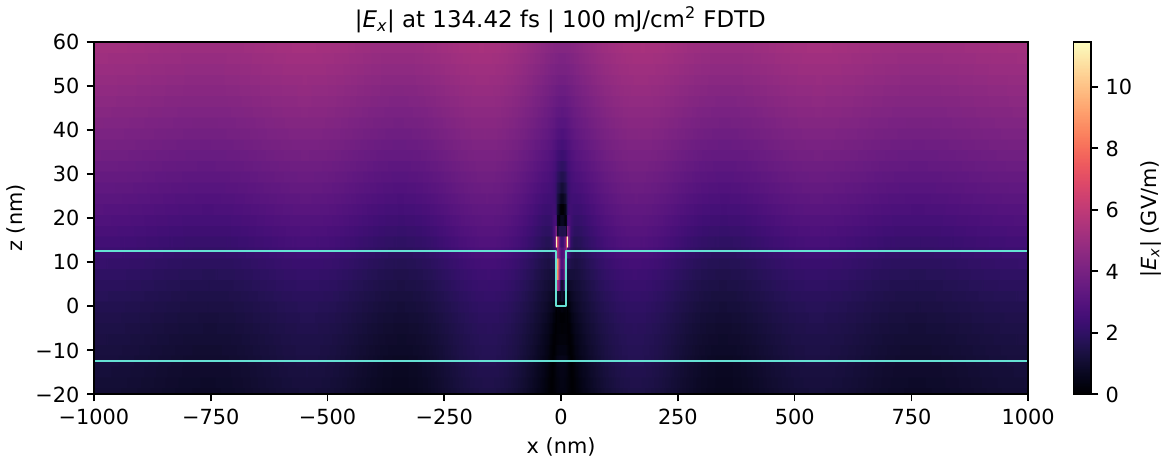}
\caption{Instantaneous magnitude of the total-field component $E_x$ at 134.42 fs in the 100 mJ/cm$^2$ FDTD calculation. The cyan line marks the nominal initial gold outline.}
\label{fig:groove-fields}
\end{figure}

\begin{figure}[htbp]

\centering
\includegraphics[width=\linewidth]{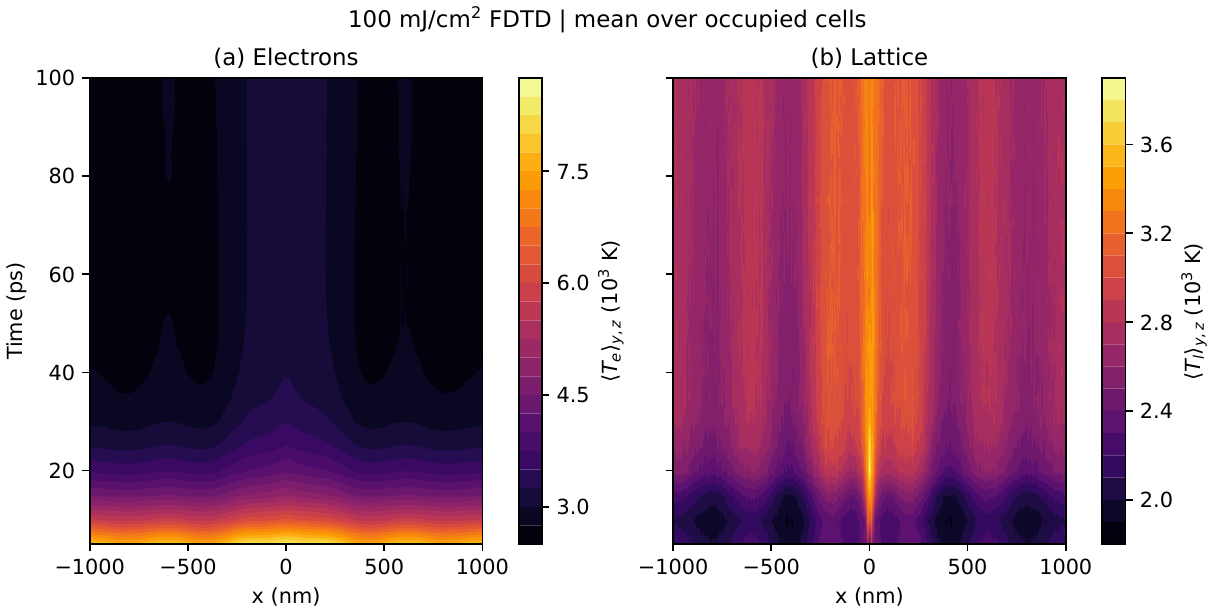}
\caption{Spatial and temporal evolution of (a) electron and (b) lattice temperatures in the 100 mJ/cm$^2$ FDTD run. At each $x$, values are arithmetic means over occupied cells in $y$ and $z$, with vacuum excluded. Filled contours connect spatial snapshots saved every 5 ps from 5 to 100 ps. The panels use separate temperature scales; the femtosecond heating peak lies outside the plotted interval.}
\label{fig:groove-temperatures}
\end{figure}

We check energy conservation in Fig.~\ref{fig:groove-energy} using the balances defined in Sec.~\ref{sec:energy-balance}. Following optical absorption, electron--lattice coupling transfers energy from the electrons to the atoms, increasing their kinetic and potential energies. The accounted total remains approximately constant through 100 ps, with a maximum matter-energy-balance error below 0.1\% of the absorbed energy across the four runs.

\begin{figure}[htbp]

\centering
\includegraphics[width=\linewidth]{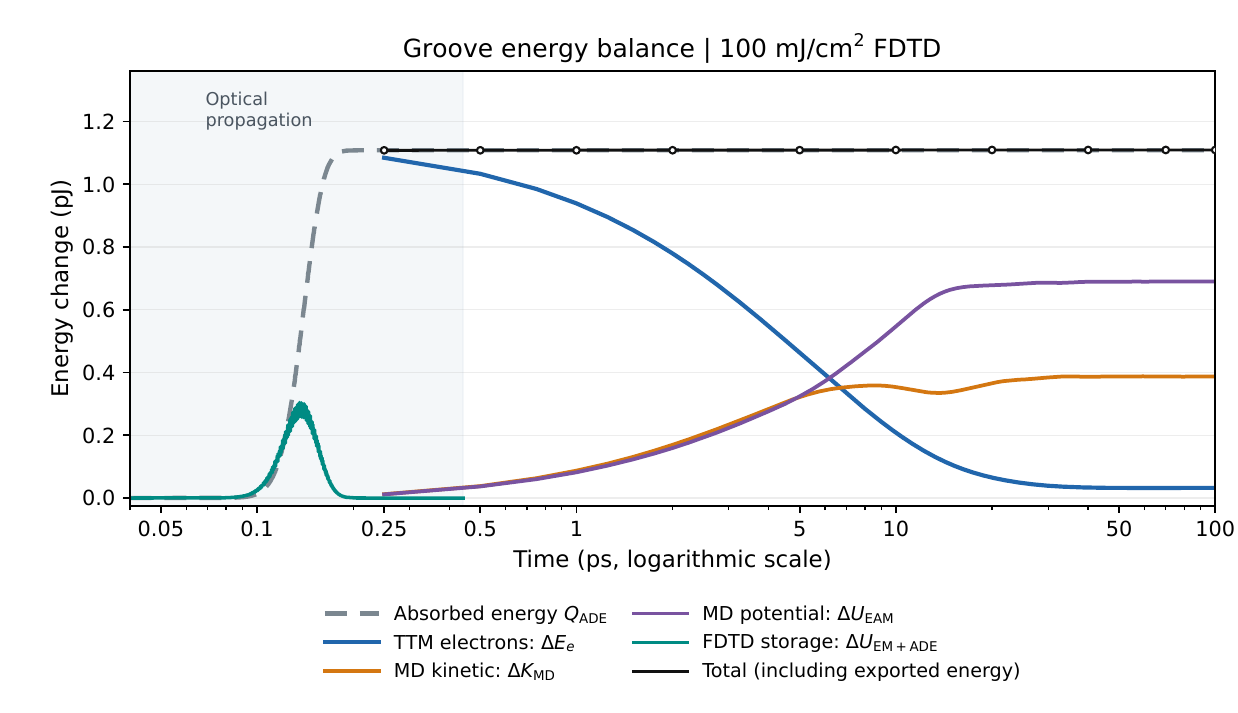}
\caption{Energy redistribution and conservation in the 100 mJ/cm$^2$ FDTD groove calculation. Energy changes are referenced to the first MD output at 2.5 fs. The black curve sums the electron, atomic kinetic, atomic potential, and stored optical energies, including the recorded energy exports. The dashed absorbed-energy curve is a comparison reference and is not added to this sum. Optical quantities use their native output times; the TTM, MD, and total curves begin at 0.25 ps and use outputs spaced by 0.25 ps. The time axis is logarithmic to display the optical transient and the subsequent relaxation together.}
\label{fig:groove-energy}
\end{figure}

\FloatBarrier

\subsection{Optical response of a rough gold surface}\label{sec:euv-comparison}

In this section, we examine how the wavelength changes the electric-field distribution above the same rough gold surface. We perform FDTD calculations at 20, 30, and 40 nm to compare the spatial variations produced by the interaction of the incident light with the surface roughness. We use an MD-relaxed film with an approximately $200\times200$ nm lateral period and a gold volume equivalent to a 30 nm thick flat film. Figure~\ref{fig:euv-lateral}(a) shows the initial surface, whose RMS height variation is approximately 0.82 nm. We hold the atomic geometry fixed during optical propagation and apply periodic boundaries in $x$ and $y$, with CPMLs along $z$.

We use normally incident pulses with a 5 fs intensity FWHM and an incident fluence of 0.1 mJ/cm$^2$, calibrated using matched vacuum calculations. At each wavelength, we fit a passive Drude response to the corresponding cold CXRO permittivity \cite{cxroAu2026}. The field maps shown here use a grid spacing of approximately 1 nm.

To compare the wavelengths on the same basis, we sample the cycle-averaged electric-field intensity at the same time and on the same horizontal plane above the entire surface. In Figs.~\ref{fig:euv-lateral}(b)--(d), we show the percentage difference between the local intensity and its average over that plane. Positive values therefore identify regions where the intensity is greater than the plane average, while negative values identify regions where it is lower. We normalize each wavelength by its own plane average and use the same color scale for all three maps.

\begin{figure}[p]

\centering
\includegraphics[width=\linewidth]{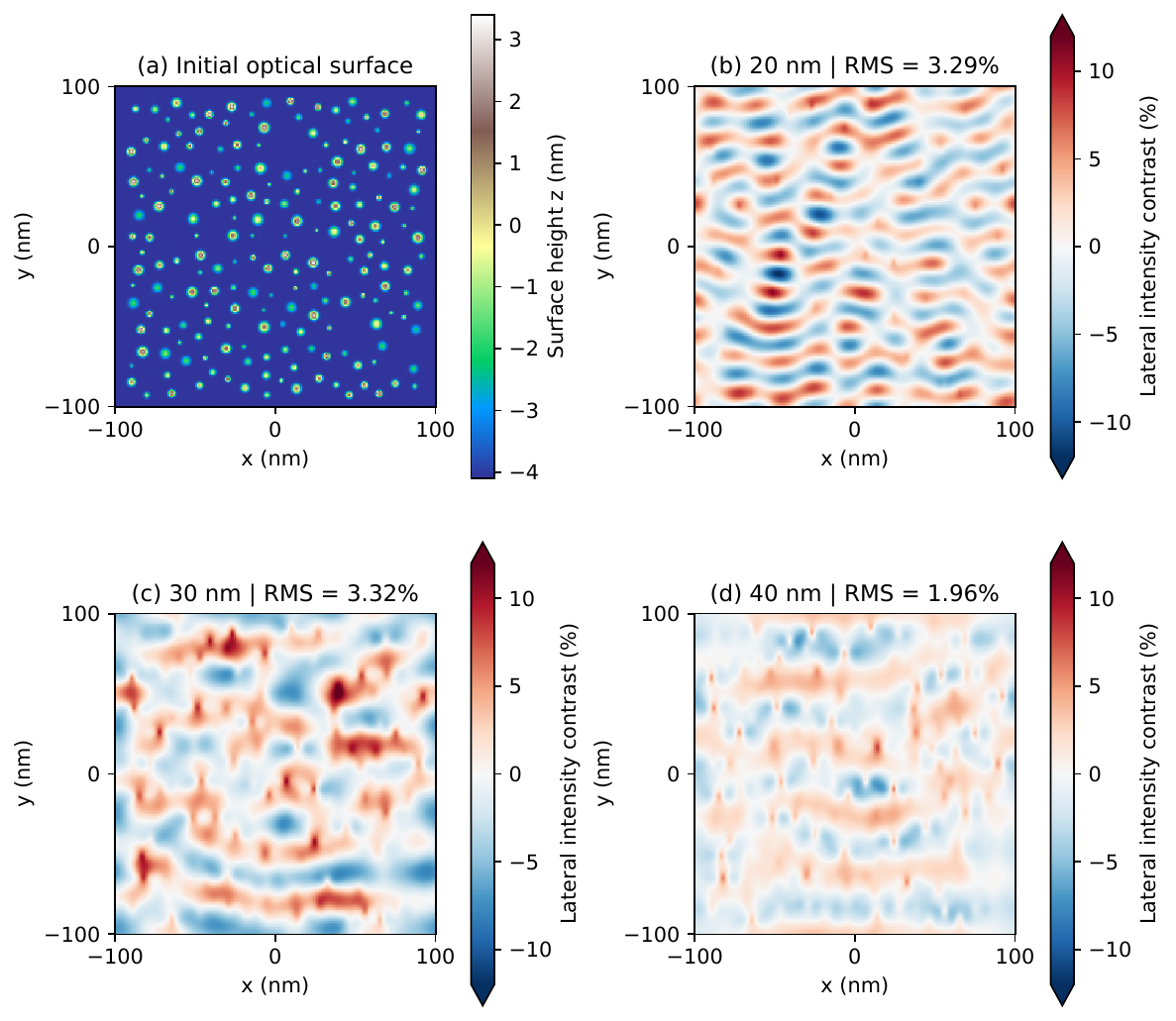}
\caption{Initial rough surface and spatial variations in the electric-field intensity above it. (a) Optical surface before irradiation. (b)--(d) Percentage deviation of the local, cycle-averaged intensity from its average over the observation plane, sampled at $z=6$ nm and $t=17$ fs. Positive values indicate stronger intensity than the plane average, and negative values indicate weaker intensity. All maps use the same color scale. The RMS values summarize the strength of the spatial variations.}
\label{fig:euv-lateral}
\end{figure}

We observe stronger spatial variations at 20 and 30 nm than at 40 nm. Surface roughness scatters the incident light, and interference between the incident and scattered fields creates regions of stronger and weaker intensity above the surface. Changing the wavelength alters this pattern through both the optical response of gold and the interaction of the light with the surface features. These calculations reveal qualitative differences in the spatial intensity distribution above the same rough surface at different wavelengths. Our independent integrated energy check gives a deviation of approximately 0.4--1.5\% of the incident energy.

\FloatBarrier

}

\section{Conclusions}\label{conclusion}
We have presented a further upgrade to the TTM-MD code by integrating an FDTD solver for Maxwell’s equations. The modules are coupled bidirectionally, enabling feedback between field-resolved absorption and atomistic response. By resolving the absorption landscape self-consistently and transferring it to atomistic mechanics, the model explains how femtosecond excitation seeds spatially organized morphology in metals. These results demonstrate that the framework is {physically relevant} in investigations of laser-induced patterning in metals, offering deeper understanding of spatially and temporally resolved fields, temperatures, and stresses induced by external excitations. {The added groove and rough-surface examples illustrate how the optical source affects material evolution and how the wavelength changes the fields above a rough surface.}

As demonstrated, the combined M-TTM-MD method requires substantial computational resources, drawing on both TTM-MD and FDTD components to simulate intricate light-material interactions. The simulations demonstrate how field-resolved absorption changes energy deposition and the resulting structural response, with consistent energy transfer between the coupled subsystems. While this integrated approach is able to provide insights with atomistic resolution, there is potential for optimizing the model's efficiency. One notable improvement would be implementing a multi-resolution FDTD grid, where finer resolution is used only in critical regions (such as near the material’s surface or around dense areas with tiny geometrical features) while a coarser resolution is applied elsewhere. This approach could significantly reduce the computational load and memory usage {subject to spatial and temporal convergence checks}.

To further enhance computational efficiency, one could further develop adaptive time-stepping in both the TTM-MD and FDTD components of the algorithm since it would allow for smaller time steps only when necessary, improving the overall speed of the simulation. Alternatively, using unconditionally stable methods could {} {relax stability restrictions}{} on timestep size, as discussed for FDTD techniques in \cite{taflove1993computational}. 

This version of the code represents the first iteration of many, with the aim of developing, with the proposed enhancements, a more efficient and more versatile scheme that can be used in different fields.

\section*{Supplementary material}
No supplementary material is provided.

\section*{Acknowledgements}
We wish to acknowledge the financial support of the Deutsche Forschungsgemeinschaft for the projects GA 465/27-1, GA 465/31-1 and GA 465/32-1. The authors also gratefully acknowledge the computing time provided to them on the high-performance computer Lichtenberg II at TU Darmstadt, funded by the German Federal Ministry of Education and Research (BMBF) and the State of Hesse.

\section*{Author declarations}

\subsection*{Declaration of competing interest}
The authors have no conflicts to disclose.

\subsection*{Ethics approval}
Ethics approval not required.

\subsection*{CRediT author statement}
O. Benhayoun: Conceptualization, Methodology, Software, Validation, Formal analysis, Investigation, Data curation, Visualization, Writing -- original draft. M. E. Garcia: Conceptualization, Supervision, Resources, Funding acquisition, Writing -- review \& editing.

\section*{Data availability}
Raw data were generated on the Lichtenberg II high-performance computing cluster at TU Darmstadt. Owing to the data volume and storage constraints, the full raw simulation outputs are not hosted in a public repository. Derived datasets and post-processing scripts that support the findings of this study are available from the corresponding author upon reasonable request.

\setlength{\bibsep}{0pt}
\bibliographystyle{elsarticle-num}
  \bibliography{mybibfile,revision_references}

@article{anisimov1974electron,
  title={Electron emission from metal surfaces exposed to ultrashort laser pulses},
  author={Anisimov, SI and Kapeliovich, BL and Perelman, TL},
  journal={Sov. Phys. JETP},
  volume={39},
  number={2},
  pages={375--377},
  year={1974}
}

@article{rethfeld2017modelling,
  title={Modelling ultrafast laser ablation},
  author={Rethfeld, B. and Ivanov, D. S. and Garcia, M. E. and Anisimov, S. I.},
  journal={J. Phys. D Appl. Phys.},
  volume={50},
  number={193001},
  doi={10.1088/1361-6463/50/19/193001},
  year={2017}
}

@article{yee1966numerical,
  title={Numerical solution of initial boundary value problems involving Maxwell's equations in isotropic media},
  author={Yee, Kane},
  journal={IEEE Transactions on antennas and propagation},
  volume={14},
  number={3},
  pages={302--307},
  year={1966},
  publisher={IEEE}
}

@article{ivanov2003combined,
  title={Combined atomistic-continuum modeling of short-pulse laser melting and disintegration of metal films},
  author={Ivanov, Dmitriy S and Zhigilei, Leonid V},
  journal={Physical Review B},
  volume={68},
  number={6},
  pages={064114},
  year={2003},
  publisher={APS}
}

@article{singh2010two,
  title={Two-temperature model of nonequilibrium electron relaxation: A review},
  author={Singh, Navinder},
  journal={Int. J. Mod. Phys. B},
  volume={24},
  number={09},
  pages={1141--1158},
  year={2010},
  doi={10.1142/S0217979210055366},
  publisher={World Scientific}
}

@book{taflove1993computational,
  title        = {Computational Electrodynamics: The Finite-Difference Time-Domain Method},
  author       = {Taflove, Allen and Hagness, Susan C.},
  publisher    = {Artech House},
  address      = {Boston, MA},
  edition      = {3},
  year         = {2005}
}

@article{roden2000convolution,
  title={Convolution PML (CPML): An efficient FDTD implementation of the CFS--PML for arbitrary media},
  author={Roden, J Alan and Gedney, Stephen D},
  journal={Microwave and optical technology letters},
  volume={27},
  number={5},
  pages={334--339},
  year={2000},
  publisher={Wiley Online Library}
}

@article{blumenstein2020formation,
  title={Formation of periodic nanoridge patterns by ultrashort single pulse UV laser irradiation of gold},
  author={Blumenstein, Andreas and Garcia, Martin E and Rethfeld, Baerbel and Simon, Peter and Ihlemann, J{\"u}rgen and Ivanov, Dmitry S},
  journal={Nanomaterials},
  volume={10},
  number={10},
  pages={1998},
  year={2020},
  publisher={MDPI}
}

@article{ivanov2019numerical,
  title={Numerical investigation of ultrashort laser-ablative synthesis of metal nanoparticles in liquids using the atomistic-continuum model},
  author={Ivanov, Dmitry S and Izgin, Thomas and Maiorov, Alexey N and Veiko, Vadim P and Rethfeld, Baerbel and Dombrovska, Yaroslava I and Garcia, Martin E and Zavestovskaya, Irina N and Klimentov, Sergey M and Kabashin, Andrei V},
  journal={Molecules},
  volume={25},
  number={1},
  pages={67},
  year={2019},
  publisher={MDPI}
}

@article{zhukhovitskii2020thermodynamics,
  title={Thermodynamics and the structure of clusters in the dense Au vapor from molecular dynamics simulation},
  author={Zhukhovitskii, D. I. and Zhakhovsky, V. V.},
  journal={The Journal of Chemical Physics},
  volume={152},
  number={22},
  pages={224705},
  doi={10.1063/5.0010156},
  year={2020},
  publisher={AIP Publishing}
}

@article{birnbaum1965semiconductor,
  title={Semiconductor surface damage produced by ruby lasers},
  author={Birnbaum, Milton},
  journal={Journal of Applied Physics},
  volume={36},
  number={11},
  pages={3688--3689},
  year={1965},
  publisher={American Institute of Physics}
}

@article{gurevich2016mechanisms,
  title={Mechanisms of femtosecond LIPSS formation induced by periodic surface temperature modulation},
  author={Gurevich, Evgeny L},
  journal={Applied Surface Science},
  volume={374},
  pages={56--60},
  year={2016},
  publisher={Elsevier}
}

@article{terekhin2022key,
  title={Key role of surface plasmon polaritons in generation of periodic surface structures following single-pulse laser irradiation of a gold step edge},
  author={Terekhin, Pavel N and Oltmanns, Jens and Blumenstein, Andreas and Ivanov, Dmitry S and Kleinwort, Frederick and Garcia, Martin E and Rethfeld, Baerbel and Ihlemann, J{\"u}rgen and Simon, Peter},
  journal={Nanophotonics},
  volume={11},
  number={2},
  pages={359--367},
  year={2022},
  publisher={De Gruyter}
}

@article{bonse2020maxwell,
  title={Maxwell meets Marangoni—a review of theories on laser-induced periodic surface structures},
  author={Bonse, J{\"o}rn and Gr{\"a}f, Stephan},
  journal={Laser \& Photonics Reviews},
  volume={14},
  number={10},
  pages={2000215},
  year={2020},
  doi={10.1002/lpor.202000215},
  publisher={Wiley Online Library}
}

@article{sipe1983laser,
  title={Laser-induced periodic surface structure. I. Theory},
  author={Sipe, JE and Young, Jeff F and Preston, JS and Van Driel, HM},
  journal={Physical Review B},
  volume={27},
  number={2},
  pages={1141},
  year={1983},
  publisher={APS}
}

@article{mastellone2022lipss,
  title={LIPSS applied to wide bandgap semiconductors and dielectrics: Assessment and future perspectives},
  author={Mastellone, Matteo and Pace, Maria Lucia and Curcio, Mariangela and Caggiano, Nicola and De Bonis, Angela and Teghil, Roberto and Dolce, Patrizia and Mollica, Donato and Orlando, Stefano and Santagata, Antonio and others},
  journal={Materials},
  volume={15},
  number={4},
  pages={1378},
  year={2022},
  publisher={MDPI}
}

@article{schwarz2018homogeneous,
  title={Homogeneous low spatial frequency LIPSS on dielectric materials generated by beam-shaped femtosecond pulsed laser irradiation},
  author={Schwarz, Simon and Rung, Stefan and Esen, Cemal and Hellmann, Ralf},
  journal={Journal of Laser Micro Nanoengineering},
  volume={13},
  number={2},
  pages={90--94},
  year={2018},
  publisher={Reza Netsu Kako Kenkyukai}
}

@PREAMBLE{
 "\providecommand{\noopsort}[1]{}" 
 # "\providecommand{\singleletter}[1]{#1}%" 
}

@article{bonse2009role,
  title={On the role of surface plasmon polaritons in the formation of laser-induced periodic surface structures upon irradiation of silicon by femtosecond-laser pulses},
  author={Bonse, J{\"o}rn and Rosenfeld, Arkadi and Kr{\"u}ger, J{\"o}rg},
  journal={J. Appl. Phys},
  volume={106},
  number={10},
  pages={104910},
  year={2009},
  doi={10.1063/1.3261734},
  publisher={AIP}
}

@inproceedings{anisimov1997theory,
  title={Theory of ultrashort laser pulse interaction with a metal},
  author={Anisimov, S. I. and Rethfeld, B.},
  editor={Konov, V. I. and Libenson, M. N.},
  booktitle={Nonresonant Laser-Matter Interaction (NLMI-9)},
  series={Proceedings of SPIE},
  volume={3093},
  pages={192--203},
  year={1997},
  doi={10.1117/12.271674},
  publisher={SPIE}
}

@article{terekhin2019influence,
  title={Influence of surface plasmon polaritons on laser energy absorption and structuring of surfaces},
  author={Terekhin, PN and Benhayoun, O and Weber, ST and Ivanov, DS and Garcia, ME and Rethfeld, B},
  journal={Appl. Surf. Sci.},
  pages={144420},
volume={512},
  year={2020},
  doi={10.1016/j.apsusc.2019.144420},
  publisher={Elsevier}
}

@article{ndione2024adaptive,
  title={Adaptive model for the optical properties of excited gold},
  author={Ndione, PD and Weber, ST and Gericke, DO and Rethfeld, B},
  journal={Physical Review B},
  volume={109},
  number={11},
  pages={115148},
  year={2024},
  publisher={APS}
}

@article{benhayoun2021theory,
  title={Theory for heating of metals assisted by surface plasmon polaritons},
  author={Benhayoun, Othmane and Terekhin, PN and Ivanov, DS and Rethfeld, B{\"a}rbel and Garcia, ME},
  journal={Applied Surface Science},
  volume={569},
  pages={150427},
  year={2021},
  publisher={Elsevier}
}

@article{gamaly2013physics,
  title={Physics of ultra-short laser interaction with matter: From phonon excitation to ultimate transformations},
  author={Gamaly, Eugene G and Rode, Andrei V},
  journal={Progress in Quantum Electronics},
  volume={37},
  number={5},
  pages={215--323},
  year={2013},
  publisher={Elsevier}
}

@article{gamaly2011physics,
  title={The physics of ultra-short laser interaction with solids at non-relativistic intensities},
  author={Gamaly, Eugene G},
  journal={Physics Reports},
  volume={508},
  number={4-5},
  pages={91--243},
  year={2011},
  publisher={Elsevier}
}

@article{lin2008electron,
  title={Electron-phonon coupling and electron heat capacity of metals under conditions of strong electron-phonon nonequilibrium},
  author={Lin, Zhibin and Zhigilei, Leonid V and Celli, Vittorio},
  journal={Physical Review B},
  volume={77},
  number={7},
  pages={075133},
  doi={10.1103/PhysRevB.77.075133},
  year={2008},
  publisher={APS}
}

@misc{schneider2010understanding,
  title={Understanding the Finite-Difference Time-Domain Method},
  author={Schneider, John B.},
  howpublished={Washington State University},
  year={2010},
  url={https://eecs.wsu.edu/~schneidj/ufdtd/}
}

@inproceedings{su2004novel,
  title={A novel FDTD application featuring OpenMP-MPI hybrid parallelization},
  author={Su, Mehmet F and El-Kady, Ihab and Bader, David A and Lin, S-Y},
  booktitle={International Conference on Parallel Processing, 2004. ICPP 2004.},
  pages={373--379},
  year={2004},
  organization={IEEE}
}

@article{nunes2011electromagnetic,
  title={Electromagnetic energy density in dispersive and dissipative media},
  author={Nunes, Frederico Dias and Vasconcelos, Thiago Campos and Bezerra, Marcel and Weiner, John},
  journal={Journal of the Optical Society of America B},
  volume={28},
  number={6},
  pages={1544--1552},
  year={2011},
  publisher={Optical Society of America}
}

@misc{byrnes2020multilayer,
 author={Byrnes, Steven J.}, title={Multilayer optical calculations},
 year={2020}, howpublished={arXiv:1603.02720v5},
 doi={10.48550/arXiv.1603.02720}, url={https://arxiv.org/abs/1603.02720}}

@misc{cxroAu2026,
 author={{Center for X-Ray Optics}}, title={Atomic scattering factors: Au},
 howpublished={Lawrence Berkeley National Laboratory}, year={2026},
 url={https://henke.lbl.gov/optical_constants/sf/au.nff},
 note={Archived input retrieved 5 September 2026}}

\end{document}